\documentclass[fleqn,usenatbib]{mnras}

\usepackage{newtxtext,newtxmath}
\usepackage[T1]{fontenc}
\DeclareRobustCommand{\VAN}[3]{#2}
\let\VANthebibliography\thebibliography
\def\thebibliography{\DeclareRobustCommand{\VAN}[3]{##3}\VANthebibliography}

\usepackage{graphicx}	% Including figure files
\usepackage{amsmath}	% Advanced maths commands

\newcommand\kms{km s$^{-1}$}
\newcommand\masyr{mas yr$^{-1}$}
\newcommand\teff{$T_{eff}$}

\title[Young runaways]{Searching for young runaways across the sky}

\author[M. Kounkel et al.]{
Marina Kounkel,$^{1}$\thanks{E-mail: marina.kounkel@vanderbilt.edu}
Aidan Mcbride,$^{2}$\thanks{E-mail: aidan.mcbride@utah.edu}
Keivan G.\ Stassun$^{1}$
and Nathan Leigh$^{3,4}$
\\
$^{1}$Department of Physics and Astronomy, Vanderbilt University, VU Station 1807, Nashville, TN 37235, USA\\
$^{2}$Department of Physics and Astronomy, University of Utah, JFB, 115 1400 E Salt Lake City, UT 84112, USA\\
$^{3}$Departamento de Astronom\'{i}a, Facultad de Ciencias F\'{i}sicas y Matem\'{a}ticas, Universidad de Concepci\'{o}n, Concepci\'{o}n, Chile\\
$^{4}$Department of Astrophysics, American Museum of Natural History, New York, NY 10024, USA
}

\date{Accepted XXX. Received YYY; in original form ZZZ}

\pubyear{2022}

\begin{document}
\label{firstpage}
\pagerange{\pageref{firstpage}--\pageref{lastpage}}
\maketitle

\begin{abstract}
We present a catalog of 3354 candidate young stars within 500 pc that appear to have been ejected from their parent associations with relative speeds of $>$5 \kms. These candidates have been homogeneously selected through performing a 2d spherical traceback of previously identified pre-main sequence candidates to various star forming regions, ensuring that the traceback age as well as the estimated age of a star is consistent with the age of the population, and excluding contaminants from the nearby moving groups that follow the dominant velocity currents in the field. Among the identified candidates we identify a number of pairs that appear to have interacted in the process of the ejection, these pairs have similar traceback time, and their trajectory appears to be diametrically opposite from each other, or they have formed a wide binary in the process. As the selection of these candidates is performed solely in 2d, spectral follow-up is necessary for their eventual confirmation. Unfortunately, recently released Gaia DR3 radial velocities appear to be unsuitable for characterizing the kinematics of low mass stars with ages $<$100 Myr, as the accretion, activity, and a variety of other spectral features that make them distinct from the more evolved stars do not appear to have been accurately accounted for in the data, resulting in significant artificially inflated scatter in their RV distribution.
\end{abstract}

\begin{keywords}
stars: kinematics and dynamics -- stars: pre-main-sequence -- proper motions
\end{keywords}

%%%%%%%%%%%%%%%%%%%%%%%%%%%%%%%%%%%%%%%%%%%%%%%%%%

%%%%%%%%%%%%%%%%% BODY OF PAPER %%%%%%%%%%%%%%%%%%
\section{Introduction}
Typically, young stars do not form in isolation -- they tend to form in massive molecular clouds alongside hundreds, if not thousands, of other young stars. Some of them originate in relatively diffuse environments, some are born in dense clusters. Regardless of the overall stellar density surrounding them, many of the youngest stellar objects initially are found in denser ``hubs'', consisting of $>2$ stars resembling proto-multiple systems \citep{chen2013}. Some of them will go on to form stable binaries and other higher order systems, and some will rapidly dissolve \citep{tobin2016,tobin2022}.

When three or more stars are found in close proximity to one another and they do not form a stable orbital configuration (such as a hierarchical triple), orbital energy exchange will occur \citep{leigh2013,stone2019,manwadkar2020,manwadkar2021}. Such an exchange will typically lead to an ejection of one of the stars from the system if they have finite sizes, and always if they are assumed to be point-particles. Typical ejection speeds are only a few \kms\ \citep{reipurth2010}, which would artificially broaden the overall velocity dispersion of the population in which these stars have been formed \citep{kounkel2021a}, but otherwise not be immediately apparent. The most extreme ejection events can accelerate stars to speeds of several tens of \kms, such stars are considered walkaway (if their speed is $<$30 \kms) and runaway \citep[if their speed is $>$30 \kms, e.g.,][]{schoettler2020}, though such boundaries are arbitrary, largely driven by the precision in velocity in the early studies that were needed for a robust confirmation that such stars have been ejected, particularly if a point of origin is uncertain.

One of the most famous pair of runaways are two O stars, AE Aur and $\mu$ Col, moving in opposite directions from one another, having been ejected $\sim2$ Myr ago from the Orion Nebula Cluster (ONC) with speeds of $\sim$100 \kms\ \citep{blaauw1961}. A number of other OB runaways have also been identified \citep{hoogerwerf2001}: as they are bright and fast moving, proper motions for a number of OB runaways has been readily available for several decades. And, in many cases, they can be traced back to specific clusters \citep{bhat2022} from which they are believed to originate.

However, identifying low mass ejected stars has presented a greater challenge. An unequivocal confirmation of youth of a star located outside of notable star forming regions has been difficult to come by. This is necessary to separate likely runaway candidates from ordinary field stars that happen to have similar kinematics -- as the latter stars are far more numerous, even a small degree of contamination can overwhelm the sample of runaways. Furthermore, it is only recently, with the release of astrometry from Gaia \citep{gaia-collaboration2018, gaia-collaboration2021} that the sufficiently precise proper motions have become available for a large number of stars.

\begin{figure*}
\includegraphics[width=\textwidth]{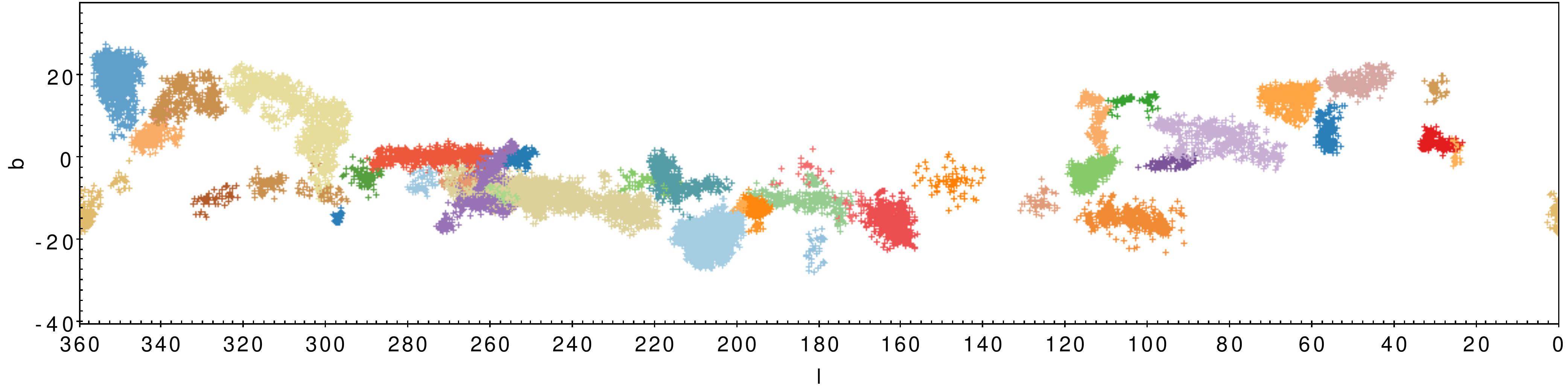}
\includegraphics[width=\textwidth]{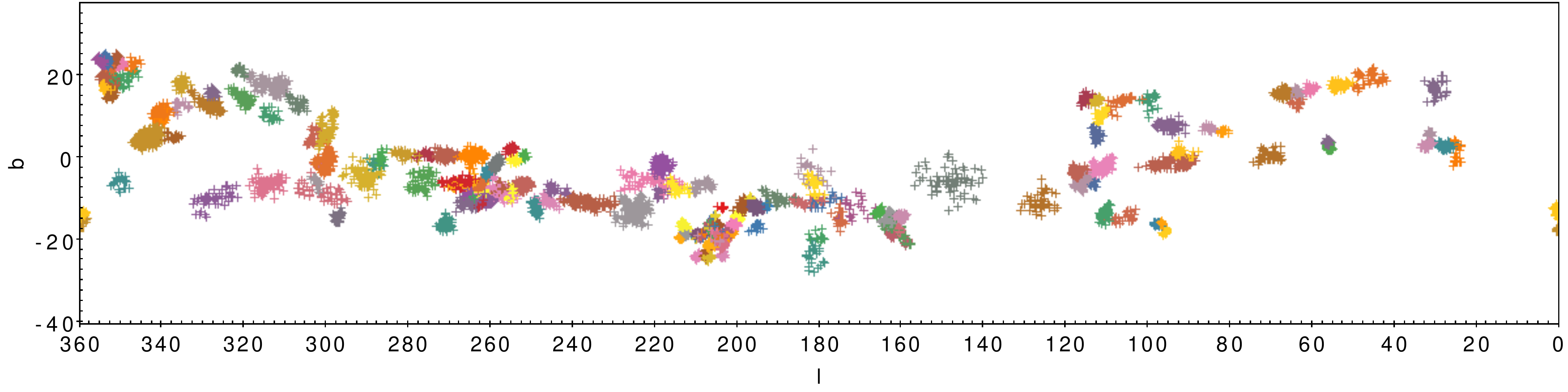}
\caption{Spatial distribution of the identified structures, arbitrarily colored by their assigned group. Top: large scale clustering. Bottom: finer subgroups.
\label{fig:clusterer}}
\end{figure*}

Since then, there have been several studies that searched for young runaway and walkaway stars, but, such studies tended to primarily focus on those that originate from the ONC. In particular, \citet{mcbride2019} have found 26 stars with tangential velocities $>$10 \kms\ among the known members of the cluster. \citet{schoettler2020} have identified 85 candidates younger than 4 Myr that trace back to the ONC, currently located within 100 pc of it. Similarly, \citet{farias2020} have identified 17,000 candidates within 45$^\circ$ of the cluster that are expected to be young -- while most of these sources suffer from contamination, 25 candidates have a particularly high likelihood of having been ejected. Other studies that have identified runaway/walkaway stars in the ONC include \citet{kounkel2017,platais2020,maiz-apellaniz2021a}.

Outside of the ONC, no systematic search of young ejected stars has been performed, and all the candidates so far are serendipitous, such as, e.g., \citet{luhman2018} identifying a dissolution of three stars in Taurus, or \citet{tajiri2020} finding a fast moving dipper star.

In this paper we take advantage of newly available catalogs of likely pre-main sequence (PMS) stars to search for runaway and walkaway candidates, tracing them back to a variety of star forming regions within 500 pc. In Section \ref{sec:data} we describe the data that form the basis of this analysis. In Section \ref{sec:methods} we present the methodology used to identify them. In Section \ref{sec:discussion} we discuss these results, and offer conclusions in Section \ref{sec:conclusions}.

\section{Data} \label{sec:data}
\subsection{Base catalog}

The release of data from Gaia has made it increasingly more possible to search for PMS stars with ages of a few to several tens of Myr. A large number of catalogs have taken advantage of the fact that young stars form in large populations that are dynamically cold to identify a large number of young moving groups \citep[e.g.,][]{kounkel2019a,kounkel2020,kerr2021,prisinzano2022}. However, as ejected stars have been dynamically processed, they may no longer share the kinematics with their parent association; as such it is necessary to confirm that a specific star is pre-main sequence without relying on its kinematics.

Other studies, such as \citet{zari2018} and \citet{mcbride2021} have performed a photometric selection of young stars along their HR diagram. In particular, \citet{mcbride2021} have developed a neural net Sagitta that was trained on the 2MASS and Gaia photometry, as well as parallaxes, of the stars that are members of young moving groups from \citet{kounkel2020}.

Sagitta consists of two parts. The first component is a classifier that assigns a probability of a particular star being pre-main sequence, separating out the young stars from the stars on the main and binary sequences, as well as from reddened high mass and red giant branch stars. Sources with the PMS probability $>0.95$ represent the cleanest sample that have only negligible contamination from the evolved stars. At lower thresholds, contamination increases, to as much as $\sim$50\% at PMS probability of $\sim0.85$; however, it becomes more sensitive towards stars with an age of a few tens of Myr, as they are closer to the binary sequence.

The second component of Sagitta is estimating the ages of the stars, using a neural net to interpolate across the empirical isochrones of the full census of members of young moving groups with known ages.

In total, \citet{mcbride2021} have identified $\sim$450,000 PMS candidates in Gaia EDR3 data down to PMS probability of 0.7 at distances of up to $\sim$3 kpc. In this work we limit the sample to only the sources with PMS$>$0.85 (to ensure that the majority of sources would be bona fide PMS stars, despite the contamination), and with parallax $\pi>2$ mas (to ensure high degree of precision in the distance and tangential velocity of stars, as well as sensitivity to low mass stars, both of which decrease substantially beyond that limit), resulting in 70,528 stars. Such cuts ensure high degree of completeness across regions included in this volume, a relatively limited degree of contamination, as well as a high precision in the Gaia astrometry across all of the candidates to ensure a reliable traceback.

\subsection{Initial clustering}

In order to identify runaways, it is necessary to evaluate their position and velocity relative to a specific population. At the same time, stars that are members of one star forming region may systematically trace back to a different star forming region due to the global dynamics -- while there may be some physical process that may relate these populations, such stars are highly unlikely to be runaways in the true sense of the word, i.e., all of their individual stars would not have been ejected through the dynamical interactions. As such, it is necessary to identify all the stars that are the co-moving members of star forming region, to exclude them from the possible pool of candidate runaways, as well as to define typical positions and velocities to which runaways can track. This is possible to do with hierarchical clustering, such as with HDBSCAN \citep{hdbscan1,hdbscan}.

However, star forming regions tend to be extended, often spanning $>$100 pc, and they may have a significant velocity gradients, which is important to take into consideration. A single position/velocity combination for an entire region (e.g., for Orion, or for Sco Cen) does not offer a sufficient degree of precision in performing such a traceback. But, similarly, if we consider positions and velocities of individual stars within a given star forming region, then this introduces too much noise in the traceback. Thus it is necessary to identify smaller but meaningful subgroups within each region.

It is possible to ``tune'' HDBSCAN to recover structures of different scales. But, unfortunately, if it is tuned to recover smaller subgroups, it would struggle to pick up stars in the more distributed parts of the star forming regions that could be easily picked up if a full population is considered. As previously mentioned, clustering is important for rejecting false positives from global kinematics, thus their exclusion is not ideal. 

Because of this, we perform a two step approach to clustering - first to identify large scale structure, and second to identify subgroups in the identified populations. This is similar to the approach considered by \citet{zari2019} and \cite{kerr2021}.

The clustering was performed in 7 dimensions:

\begin{itemize}
    \item $\arccos(\cos{(l)})/2$ deg
    \item $\arcsin(\sin{(l)})/2$ deg
    \item $b/2$ deg
    \item $\log_{10}(1000/\pi)$
    \item $4.74\mu_{l*}/\pi$
    \item $4.74\mu_{b*}/\pi$
    \item Age (Myr)$/10$ 
\end{itemize}

where $\pi$ is parallax in mas, and $\mu_{l*}$ and $\mu_{b*}$ are proper motions in $l$ and $b$, in \masyr, converted to the local standard of rest \citep{schonrich2010}, with galactic rotation subtracted \citep[see][]{kounkel2021a}. The implemented transformation converts them from proper motion to the tangential velocity space. In order to prevent splitting of groups at $l=0^\circ=360^\circ$ boundary, a circular encoding for $l$ is implemented. We note that as the sample has $\sigma_\pi/\pi<0.1$, the inversion of the parallax to get the distances is an acceptably precise approximation for the purposes of the analysis here. Additionally, while some of the scalings in the above parameters of the mismatched data units are arbitrary, they were decided on iteratively through the visual examination of the outputs to ensure a homogeneous selection of known populations.

To identify large scale structure, we used minimum cluster size of 40 stars, with the minimum sampling of 40 stars, and `eom' as the cluster selection method. This has selected 27,511 (out of 70,528) stars into 42 different populations. Afterwards, we applied HDBSCAN again on the already clustered stars, changing the minimum cluster size to 20 stars, minimum samples of 5 stars, and `leaf' as the cluster selection method. This selected 10451 stars in 167 different subgroups (Figure \ref{fig:clusterer}).

We note that these groups have a considerable degree of similarity with the structures identified in other all-sky clustering approaches \citep[e.g.,][]{kounkel2019a, kerr2021, prisinzano2022}, but there are minor differences that sometimes prevent a precise one-to-one match. Fundamentally, all clusering techniques trace out similar overdensities in the distribution of stars, but they differ in the ability of recovering particular populations that tend to be more diffuse, as well as in the ability of separating out the neighboring structures. The primary purpose of repeating the clustering compared to these previous works is to ensure the membership in these identified structures is as complete as possible relative to the base Gaia EDR3 catalog from \citet{mcbride2021} in particular, such that stars that may have been excluded in the initial selection in other works would be considered when identifying ``core'' members of all the groups. Independent clustering has also enabled us to better control the relative scales of the recovered structures.

To the best of our ability, we have attempted to cross-match all of the identified sub-groups to the structures that have been previously recognized in the literature, but it is not always possible. The membership of all of the groups is included in Table \ref{tab:full}, and their average properties are shown in Table \ref{tab:ave}.

\begin{table}
	\centering
	\caption{Catalog of clustered PMS candidates}
	\label{tab:full}
	\begin{tabular}{ccl} % four columns, alignment for each
		\hline
		Column & Unit & Description \\
		\hline
		Source &  & Gaia DR3 source id \\
		$l_*$ & deg & Galactic longitude \\
		$b_*$ & deg & Galactic latitude \\
		$\pi_*$ & mas & Gaia DR3 parallax \\
		$\sigma_{\pi,*}$ & mas & Uncertainty in parallax \\
		PMS &  & PMS probability from Sagitta \\
		$\sigma_{PMS}$ &  & Uncertainty in PMS probability \\
		$t_*$ & [yr] & (log) Age from Sagitta \\
		$\sigma_{t,*}$ & [yr] & Uncertainty in age \\
		$\mu_{l*}$ & \masyr & Corrected proper motions in l \\
		$\mu_{b*}$ & \masyr & Corrected proper motions in b \\
		Population &  & Name of the large scale population \\
		Subgroup &  & Name of finer subgroup in a larger population \\
		\hline
	\end{tabular}
\end{table}

\begin{table}
	\centering
	\caption{Identified groups and their properties}
	\label{tab:ave}
	\begin{tabular}{ccl} % four columns, alignment for each
		\hline
		Column & Unit & Description \\
		\hline
		Population &  & Name of the large scale population \\
		Subgroup &  & Name of finer subgroup in a larger population \\
		$l_g$ & deg & Average galactic longitude \\
		$\sigma_{l,g}$ & deg & Scatter in l \\
		$b_g$ & deg & Average galactic lattitude \\
		$\sigma_{b,g}$ & deg & Scatter in b \\
		$\pi_g$ & mas & Average parallax \\
		$\sigma_{\pi,g}$ & mas & Scatter in parallax \\
		$\mu_{l*,g}$ & \masyr & Average corrected proper motions in l \\
		$\sigma_{\mu l,g}$ & \masyr & Scatter in $\mu_{l*}$ \\
		$\mu_{b*,g}$ & \masyr & Average corrected proper motions in b \\
		$\sigma_{\mu b,g}$ & \masyr & Scatter in $\mu_{b*}$ \\
		$t_g$ & [yr] & Average (log) age \\
		$\sigma_{t,g} $ & [yr] & Scatter in age \\
		N &  & Number of stars \\
		\hline
	\end{tabular}
\end{table}
 
\section{Analysis}\label{sec:methods}
\subsection{Selection of candidate ejected stars}

In order to identify all of the candidate runaway and walkway stars, as well as their origin, we evaluate the position and velocity of all of the stars that were not identified as members of any of the moving groups relative to the average position and velocity of all substructures. Stars that have been ejected from a particular group would appear to be moving on a radial trajectory away from it in its rest frame.

We perform a traceback solely in the plane of the sky, similarly to \citet{farias2020} using an approximation of a spherical surface. In such a geometry, a path of a star is described by a great circle.

A bearing $\theta$ represents an angle that an object needs to travel along a great circle that would connect a starting location with a particular point on a spherical surface. For each star, we calculate the bearing $\theta$ corresponding to its proper motions via\footnote{\url{http://www.movable-type.co.uk/scripts/latlong.html}}

\begin{equation}
\begin{split}
\theta=\mathrm{atan2}(\sin(l_\mu-l_*)\cos b_\mu,
\cos b_* \sin b_\mu - \\
\sin b_* \cos b_\mu \cos (l_\mu-l_*))
\end{split}
\end{equation}

The galactic coordinates $(l_*, b_*)$ correspond to the current position of the star, and $(l_\mu, b_\mu)$ are the position of a star that are respectively offset by $\mu_{l*}$ and $\mu_{b*}$ in the reference frame of a particular group defined by that group's typical proper motions $\mu_{l*,g}$ and $\mu_{b*,g}$. I.e,.

\begin{equation}
\begin{aligned}
l_\mu &= l_*-(\mu_{l*}-\mu_{l*,g})\Delta t/ \cos(b_*)\\
b_\mu &= b_*-(\mu_{b*}-\mu_{b*,g})\Delta t
\end{aligned}
\end{equation}
\noindent with $\Delta t$ set arbitrarily small, to 1 year, and ensuring appropriate unit conversion.

We also calculate an angular distance $d$ from the star to the median position of each given group $(l_g, b_g)$ as
\begin{equation}
\phi=\mathrm{acos}(\sin b_* \sin b_g  +\cos b_* \cos b_g \cos(l_g-l_*) )
\end{equation}

Finally, we determine the initial traceback position $(l_{o}, b_{o})$ from which a star would have originated having moved along the bearing $\theta$ over the angular distance $\phi$.

\begin{equation}
\begin{split}
b_{o} =\mathrm{asin}(\sin b_* \cos \phi + \cos b_*\sin \phi \cos\theta)\\
l_{o} =l_* + \mathrm{atan2}(\sin\theta\sin \phi\cos b_*, \cos \phi-\\
\sin b_* \sin b_{o})
\end{split}
\end{equation}

We then establish a set of criteria to evaluate whether a star could have originated from a given group.

\begin{itemize}
    \item $|l_o-l_g|<3\sigma_{l,g}$, $|b_o-b_g|<3\sigma_{b,g}$, where $\sigma_{lb,g}$ is the standard deviation in $l,b$ of all the identified members of a group
    \item $\phi<50$ deg (to prevent the approximation of the great circle from breaking down at large angles)
    \item Relative velocity between the star and a moving group $v_{rel}>$5 \kms
    \item Traceback time $t_{rel}\equiv \phi/v_{rel}$ (ensuring appropriate unit conversion) is less than the age of a star ($t_*$)
    \item Traceback time is less than the typical age of a group ($t_g)$
    \item $|t_*-t_g|<$ 3 Myr, OR $|t_*-t_g|<(\sigma_{t,g})$, where $\sigma_{t,g}$ is the standard deviation of ages of all of the identified members of a group (3 Myr was chosen to ensure that very young stars would be recovered if they have been ejected from a cluster like the ONC which has sustained its star formation for a few Myr)
    \item $|1000/\pi_*-1000/\pi_g|<$100 pc, where $\pi_*$ and $\pi_g$ are the parallaxes of a star and of a group (since RVs are typically not available, this is an arbitrarily small distance that should be possible to be traversed by an ejected star in a few Myr, within errors)
    \item A star is not already a member of any of the moving groups
\end{itemize}

The catalog of all of the identified candidate runaway and walkaway stars is presented in Table \ref{tab:runaways}

\begin{table}
	\centering
	\caption{Catalog of candiate runaway and walkaway stars}
	\label{tab:runaways}
	\begin{tabular}{ccl} % four columns, alignment for each
		\hline
		Column & Unit & Description \\
		\hline
		Source &  & Gaia DR3 source id \\
		$l_*$ & deg & Galactic longitude \\
		$b_*$ & deg & Galactic latitude \\
		$\pi_*$ & mas & Gaia DR3 parallax \\
		$\sigma_{\pi,*}$ & mas & Uncertainty in parallax \\
		PMS &  & PMS probability from Sagitta \\
		$\sigma_{PMS}$ &  & Uncertainty in PMS probability \\
		$t_*$ & [yr] & (log) Age from Sagitta \\
		$\sigma_{t,*}$ & [yr] & Uncertainty in age \\
		$\mu_{l*}$ & \masyr & Corrected proper motions in l \\
		$\mu_{b*}$ & \masyr & Corrected proper motions in b \\
		Subgroup &  & Name of finer subgroup in a larger population \\
		$\theta$ & deg & Bearing of a stellar motion relative to the subgroup \\
		$\phi$ & deg & Current separation between the star and the subgroup \\
		$v_{rel}$ & \kms & Relative velocity between the star and the subgroup \\
		$t_{rel}$ & Myr & Estimated traceback time to the subgroup \\
		$l_o$ & deg & Estimated galactic longitude of the point of origin \\
		$b_o$ & deg & Estimated galactic latitude of the point of origin \\
		$\xi_1$ &  & Angle-based metric for moving group contaminants \\
		$\xi_2$ &  & Distance-based metric for moving group contaminants \\
		Wide &  & Index of wide binary candidates \\
		Opposing &  & Index of opposing pairs candidates \\
		\hline
	\end{tabular}
\end{table}

\subsection{Remaining structure identification}

The initial clustering analysis of the catalog has allowed to exclude a number of false positives runaway candidates through limiting a sample of stars that trace back to other moving groups due to dominant global kinematics in the region. However, a significant fraction of runaway candidates is still dominated by such sources. The primary source of contamination are very sparse moving groups (e.g., Taurus), comprehensive membership of which is difficult to recover with clustering in a presence of structures that are much more massive. Additionally, older groups that are in the process of dissolving have members that may be missed due to these stars having kinematics just outside of the typical velocity dispersion, or have positions just outside of where the bulk of the stars in the moving group still reside. Such stars still follow the typical dominant velocity currents of a region in which they reside, but they may not necessarily be identified among the ``core'' members of a moving group.

The ability to trace young stars in bulk to other regions may help to reveal the processes that have shaped the velocity currents in a region. However, when several stars are found in a similar direction and at a similar distance away from a particular group, such sources are unlikely to have been ejected from said group. To evaluate the probability of a given source being a viable ejected star candidate, we derive two metrics, $\xi_1$ and $\xi_2$.

\begin{equation}
\xi_1=n_{\Delta\theta<10}/n_{tot}
\end{equation}
where $n_{\Delta\theta<10}$ is the number stars with the bearing $\theta$ within 10$^\circ$ of $\theta$ of a given star, and $n_{tot}$ is the total number of stars that trace back to a given sub-group. The more concentrated the distribution of stars is originating from a particular direction, the larger $\xi_1$ becomes, signifying a possibly unaccounted moving group. If $n_{tot}<12$, then we set $n_{tot}=12$ -- this is to prevent extreme cases of e.g., if only a single star traces back to a particular sub-group, $\xi_1$ becomes 1.

\begin{equation}
\xi_2=\sigma_{\phi_\theta}/|\overline{\phi_\theta}-\phi|
\end{equation}
if $n_{\Delta\theta<10}>5$, $\phi_\theta$ are the angular distances of all stars within 10$^\circ$ of $\theta$ of a given star, otherwise, if $n_{tot}>5$, angular distances of all stars tracing back to a given group are used. Average $\overline{\phi_\theta}$ and the standard deviation $\sigma_{\phi_\theta}$ are computed using only 10--90 percentile range of the distribution of $\phi_\theta$. This tests for the commonality of angular distances that different stars trace back over: the greater the similarity, the larger $\xi_2$ becomes. If $n_{tot}\leq5$, $\xi_2$ is set to 1.

\begin{figure}
\includegraphics[width=\columnwidth]{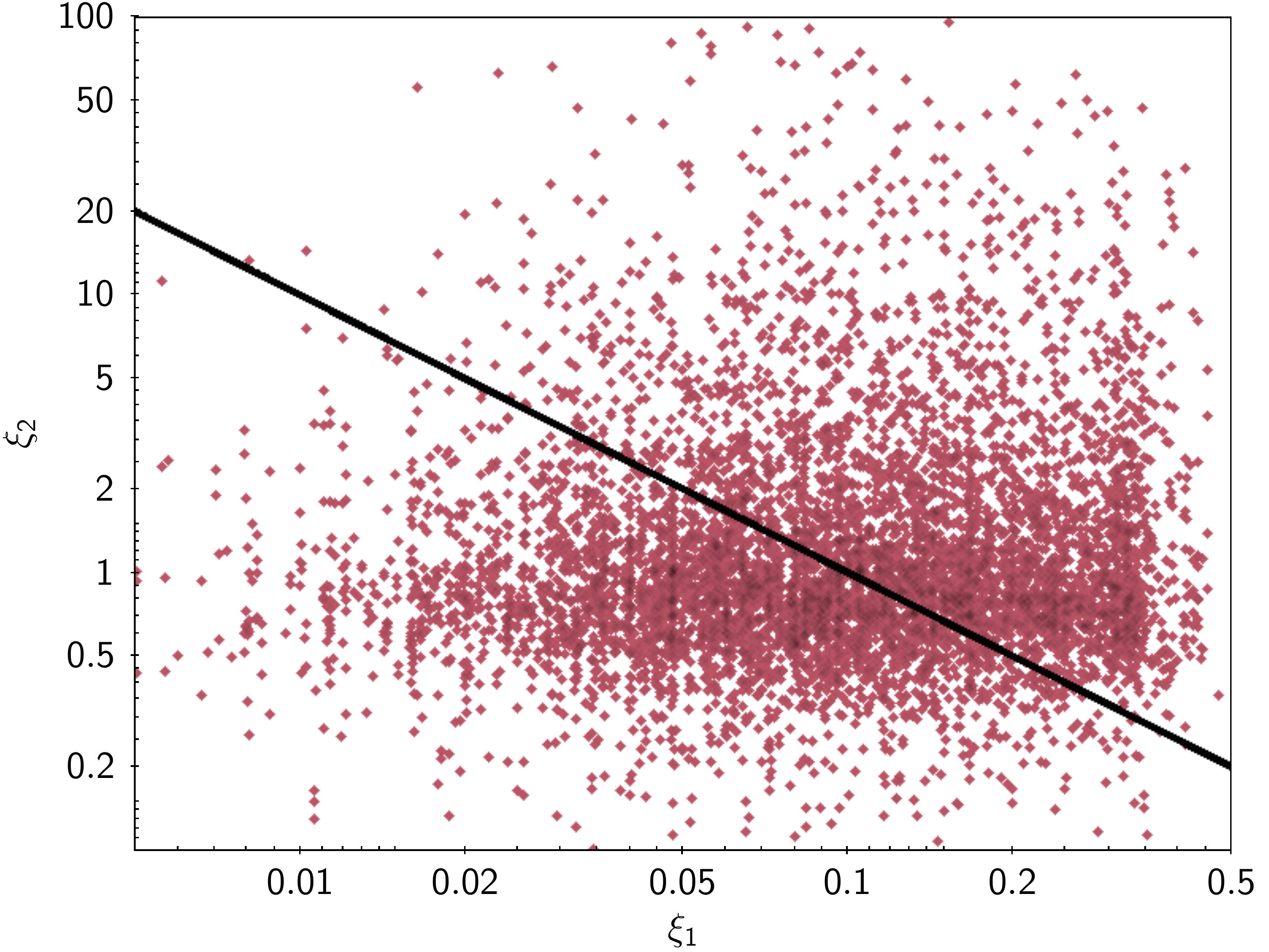}
\caption{Distribution of $\xi_1$ and $\xi_2$ parameters for all of the 6803 candidates (see text for definition). Black line corresponds to $\xi_1\xi_2<0.1$.
\label{fig:ksiplot}}
\end{figure}

\begin{figure}
\includegraphics[width=\columnwidth]{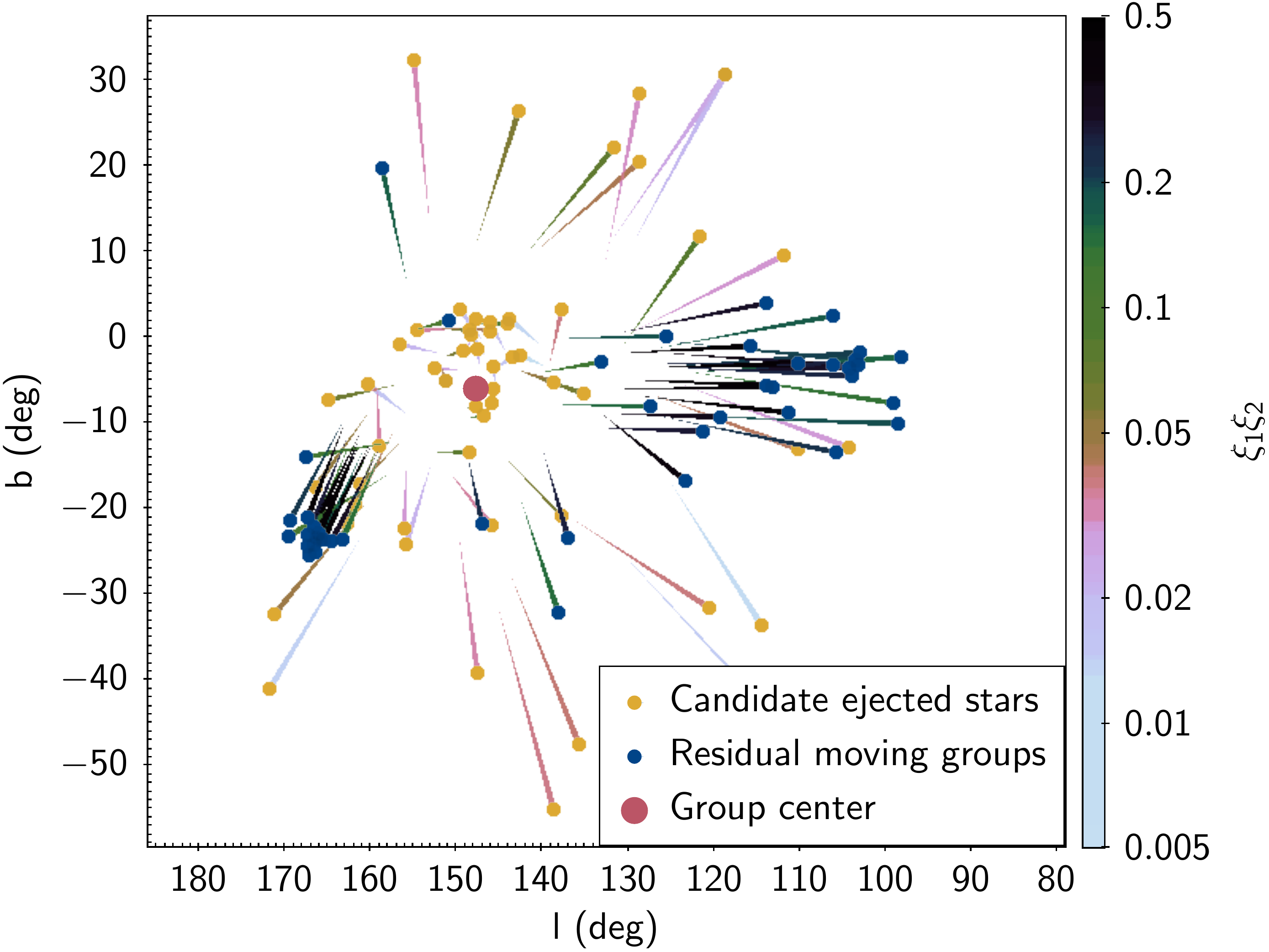}
\caption{Candidate ejected stars that trace back to one group ($\alpha$ Per), color-coded by $\xi_1\xi_2$. The sources with large $\xi_1\xi_2$ reveal the presence of two moving groups in this region that have not been recovered in the initial clustering.
\label{fig:recluster}}
\end{figure}

Conservatively, we set the threshold of $\xi_1\xi_2=0.1$ (Figure \ref{fig:ksiplot}), separating members of possibly unaccounted of moving groups and possible runaway/walkwaway stars, chosen through examining all stars tracing back to individual groups (e.g., Figure \ref{fig:recluster}), which roughly filters out all of the overdensities in the distribution of sources that are apparent to the eye.

\begin{figure*}
\includegraphics[width=\textwidth]{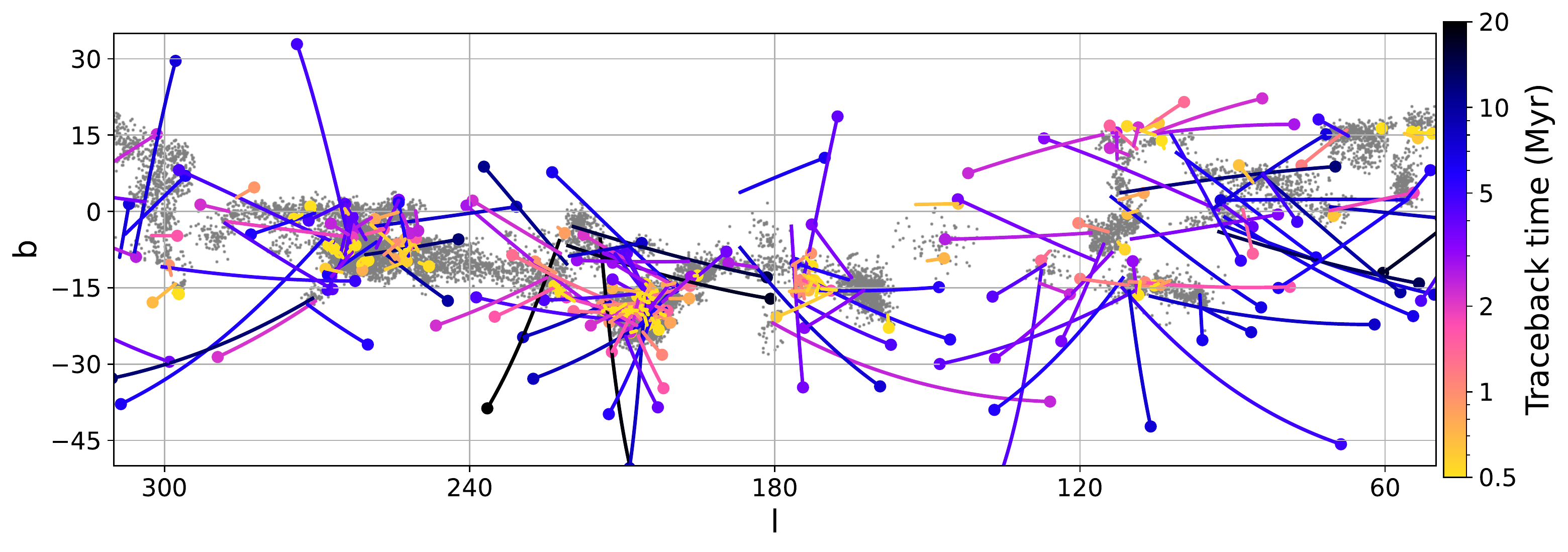}
\includegraphics[width=\textwidth]{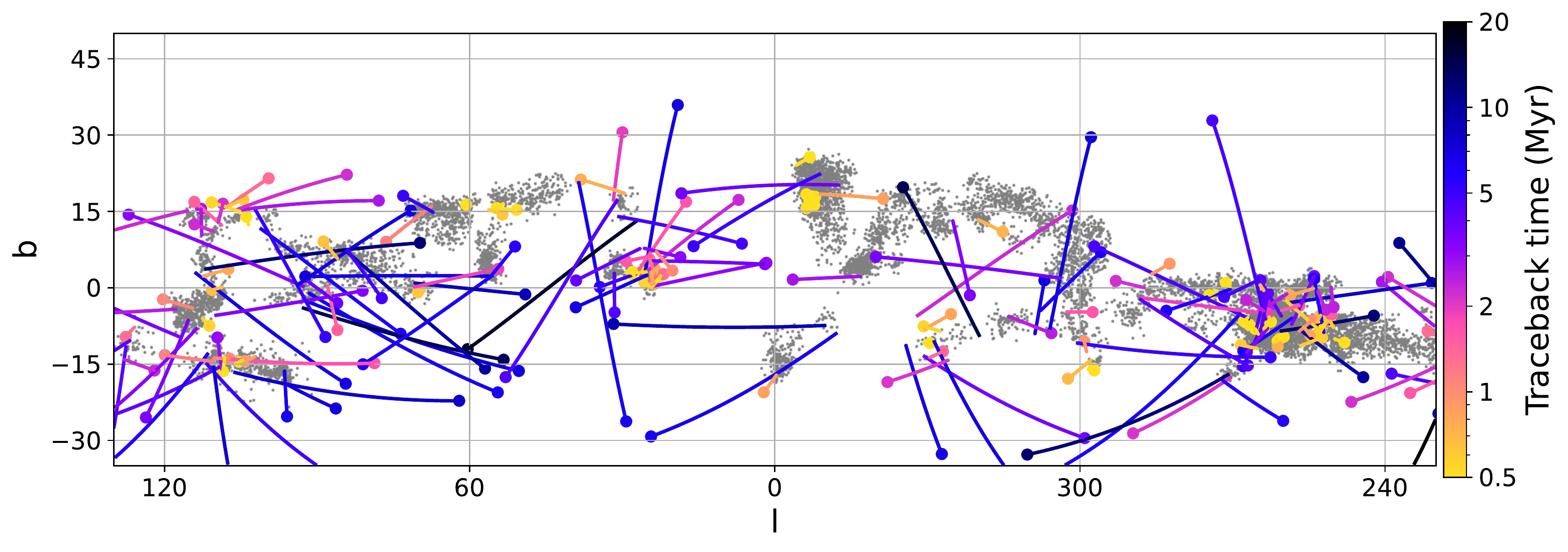}
\caption{Candidate ejected stars with PMS$>$0.95 and relative velocity$>$10 \kms. The vectors show their apparent path from the estimated $l_o$ and $b_o$ to their current position (signified by larger dot), they are color-coded by the traceback travel time. Greyscale background shows the distribution of the clustered structures.
\label{fig:runaways}}
\end{figure*}

To ensure that neighboring groups do not ``shield'' each other, preventing sources to trace towards them from a particular side, initially, a star could be matched to multiple subgroups. Following calculation of $\xi_1\xi_2$, only the group that is located the closest to a given star is chosen. In total, this selection has identified 6803 candidate runaway and walkaway stars, of which 1266 have PMS probability $>$0.95. Of these, 3449 have been excluded by $\xi_1\xi_2$ test, with only 3354 stars remaining at PMS probability $>$0.85, and only 546 at PMS probability $>0.95$ (Figure \ref{fig:runaways}).

\section{Discussion} \label{sec:discussion}

\begin{figure*}
\includegraphics[width=\columnwidth]{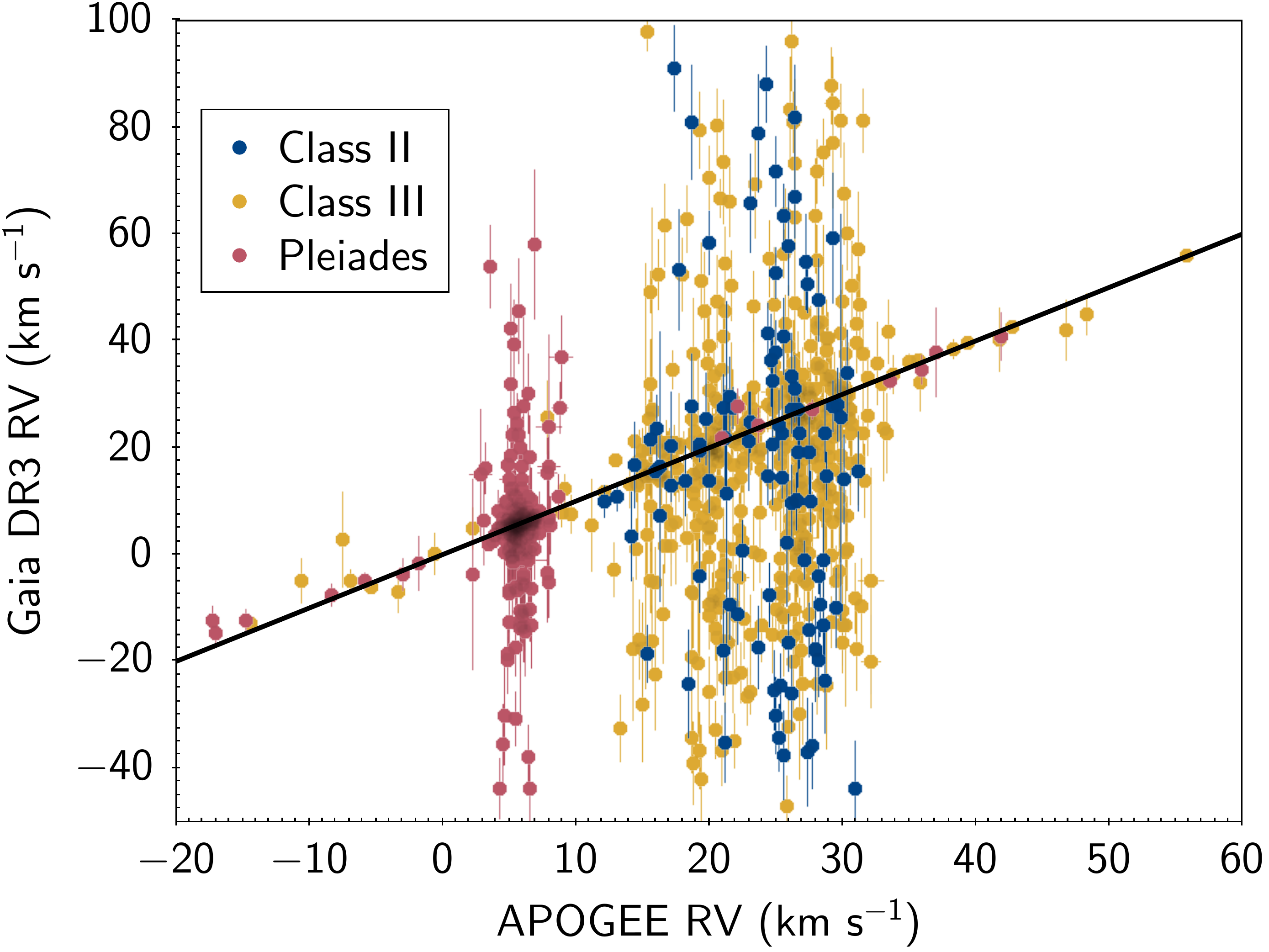}
\includegraphics[width=\columnwidth]{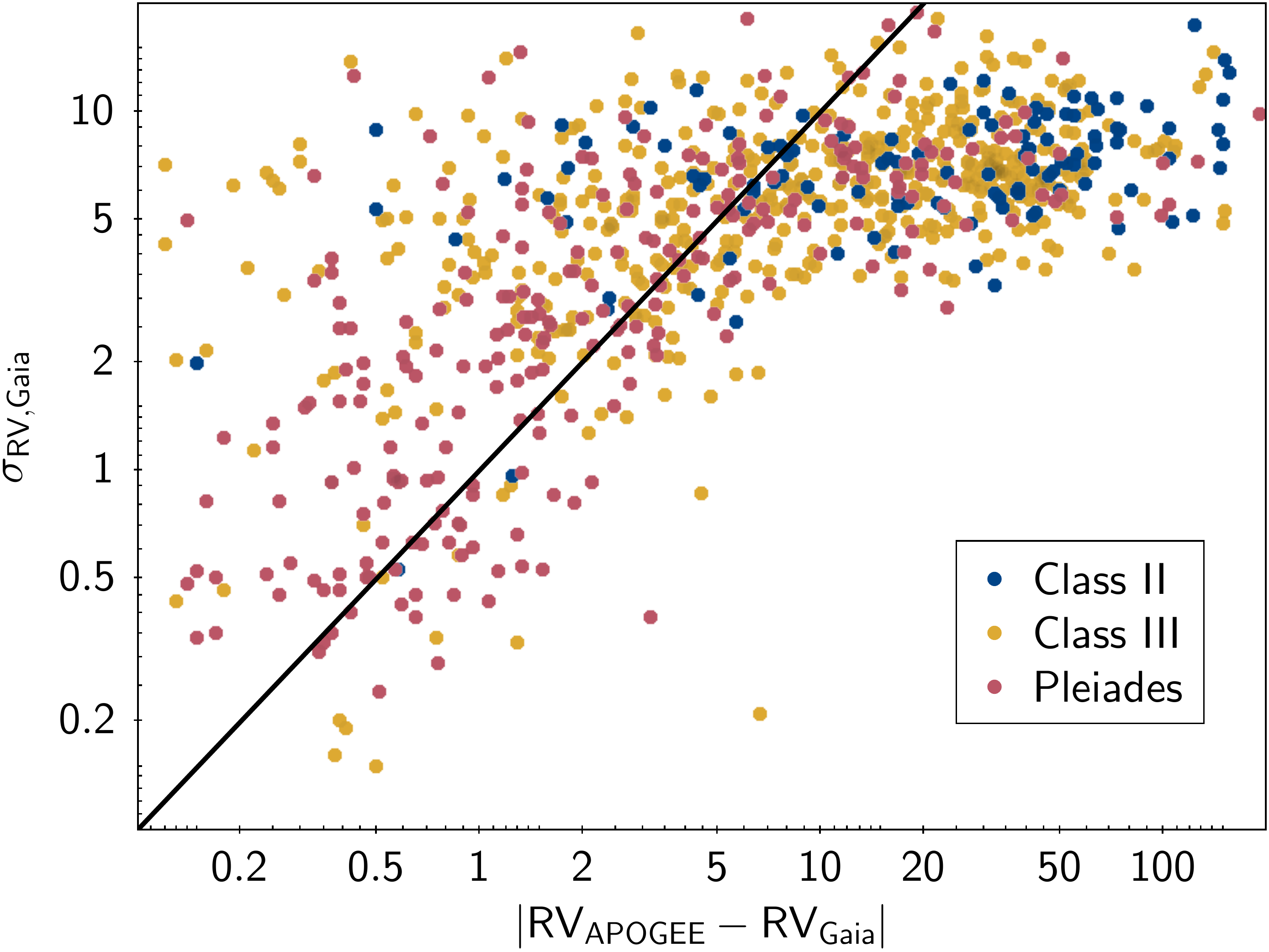}
\caption{A comparison of the Gaia DR3 and APOGEE radial velocities of young stars that appear to have RVs from \citet{kounkel2019}, from Orion, Taurus, Perseus, and NGC 2264, separated into the disk-bearing and non disk-bearing stars. A data for the Pleiades is also shown, to highlight the differences that the age of a star may create in the sample. Left panel shows the direct comparison of the two sets of RVs. Right panel shows the magnitude of the difference between the RVs as a function of the reported uncertainty in RV from Gaia DR3.
\label{fig:rv}}
\end{figure*}

\subsection{Radial Velocity}

Currently, the traceback of motion of the ejected candidates is done solely in the plane of the sky. A full 3d traceback would be able to more definitively confirm the point of origin of these fast moving pre-main sequence stars, as RVs are able to affect the overall trajectory of a star away from following a great circle.
%\textcolor{red}{NL: Perhaps for future work, but it would be super cool to see how many runaways are recovered first using 4d, then 5d, then 6d, with the same cuts applied.  I am guessing you need to full 6d info to fully and reliable catergorize a runaway.... But I do not know, and this is something Alonso will hopefully do.  If you have done some of this in thist study and feel it is reliable, I think it is worth commenting on, at least.  Lots of degeneracies in this problem...}

The RV data for PMS stars remain very sparse. The recent release of Gaia DR3 has provided RVs for over 30 million stars \citep{katz2022}, representing a major achievement in providing comprehensive data of the radial velocity structure of the Galaxy. But, the sources for which RVs are available tend to be very luminous, disfavoring low mass PMS dwarfs. Only $\sim$10\% of the base catalog of PMS stars has reported Gaia DR3 RVs, and their uncertainties tend to be 5--10 \kms. Such data are insufficient to even establish an RV reference frame for most of the populations, let alone to perform a detailed traceback of all of the individual stars fully in 3D.

Furthermore, evaluating the available Gaia RVs for PMS stars paints a somewhat grim picture regarding their quality. We compare these RVs versus those that have been measured by APOGEE in various star forming regions, such as Orion, Taurus, Perseus, and NGC 2264 \citep{kounkel2019}. We limit the sample to only those sources for which at least 3 high resolution spectroscopic observations exist to ensure RV stability, excluding any of the SB1s, SB2s (or higher order multiples) that have been identified in that work.

While the APOGEE RVs of these young stars tend to have low velocity dispersion of less than a few \kms, same sources result in the velocity dispersion of several 10s of \kms\ with Gaia RVs. This scatter is not accounted for in the uncertainties: even though the the typical reported errors in Gaia RVs are 5--10 \kms, the difference between Gaia and APOGEE RVs on the order of 10$\sigma$ is common (Figure \ref{fig:rv}).

Class II young stellar objects, i.e., sources that still have infrared excess due to the presence of a protoplanetary disk, sources that typically have strong accretion signature in optical spectrum have RVs that are qualitatively worse than in the sources that have already depleted their disk. Class III objects have poor RVs as well, however. These sources still have low surface gravity, and often may have strong activity features that may have been unaccounted in the spectral fitting. As such, Gaia RVs of nearly all young stars are unstable.

This issue persists for $>$100 Myr. For example, in Pleiades, where almost all of the stars are already on the main sequence, but many still show high activity. While the solar-type stars (\teff$\sim$6000 K) have reasonable RV precision, with the scatter in RVs reproduced by the uncertainties (typically $<$2 \kms), the cooler dwarfs in Pleiades ($\sim$1/4 of the sample) also have RVs with unrealistically large scatter. More evolved field dwarfs in the APOGEE data that are not associated with any young cluster or a moving group do not appear to be affected to a similar degree, even in comparison to the Pleiades. As such, this issue appears to affect younger and active stars specifically.

As our sample consists entirely of young cool dwarfs, we chose not to use Gaia RVs even in the cases where these sparse data are available. Furthermore, we advise extreme caution to other studies that attempt to over-interpret the observed features of the RV distribution in the young stars \citep[such as][]{zucker2022a} -- the improper data processing of Gaia spectra provides yet another source of RV scatter that needs to be considered carefully in determining such parameters as e.g., velocity dispersion of young moving groups.

Unfortunately, at the moment, no other RV survey provides a comprehensive coverage of data across all of the candidates. This will eventually change: over the next five years, SDSS-V is set to obtain high resolution spectra $\sim$100,000 young stars (Kounkel M., in prep), however, these data are not yet available. It is also possible that in the future a more careful independent processing of the spectroscopic data released by Gaia will enable a more optimal RV extraction of the young stars -- once RVS spectra from Gaia are released in full, or in the subsequent data releases.

Ultimately, a comprehensive confirmation of the candidate ejected stars, fully in 3d, may not be possible now, but such an analysis will be possible at a later date.

\subsection{Comparison to previous works}

\begin{figure*}
\includegraphics[width=\columnwidth]{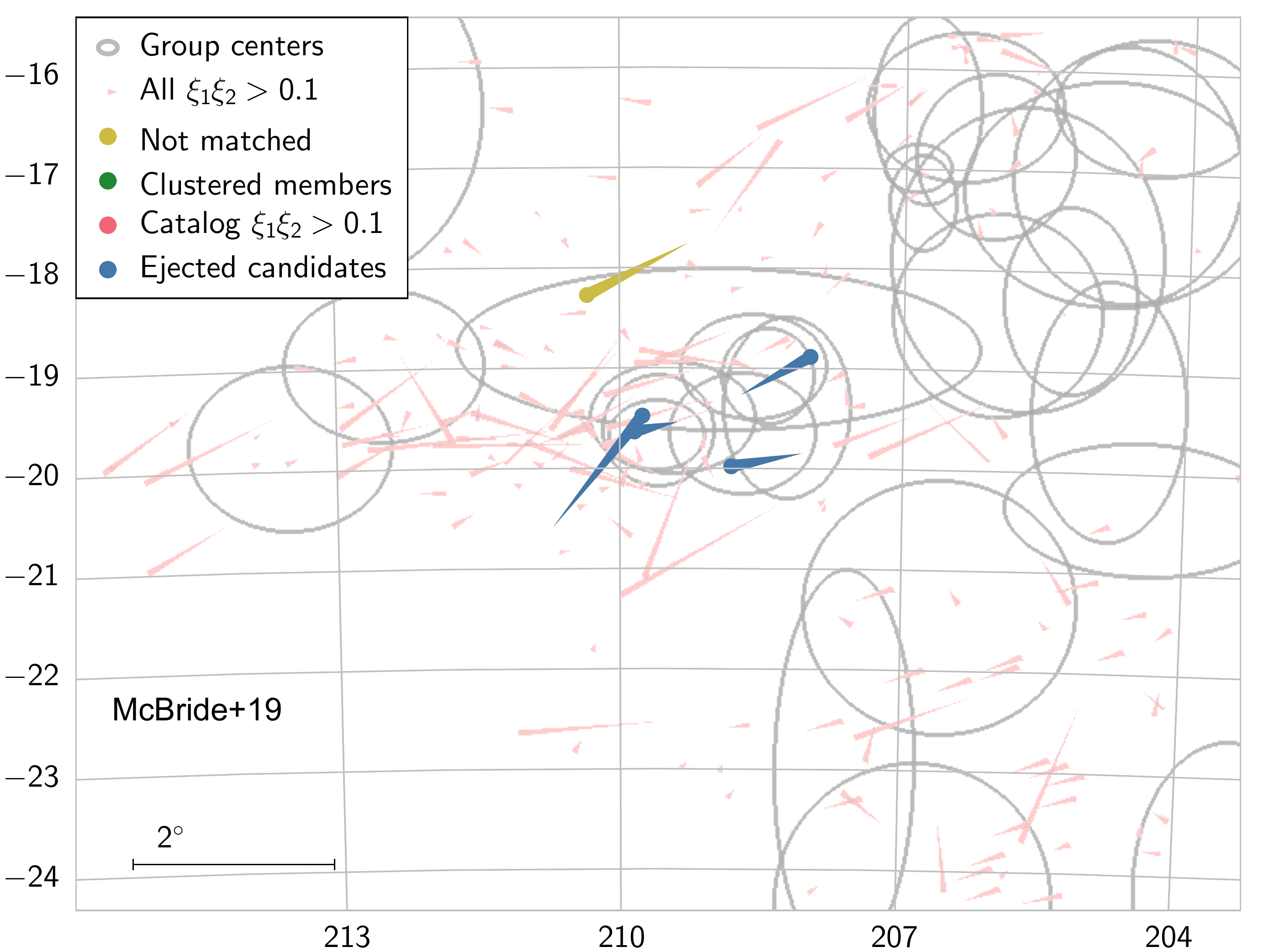}
\includegraphics[width=\columnwidth]{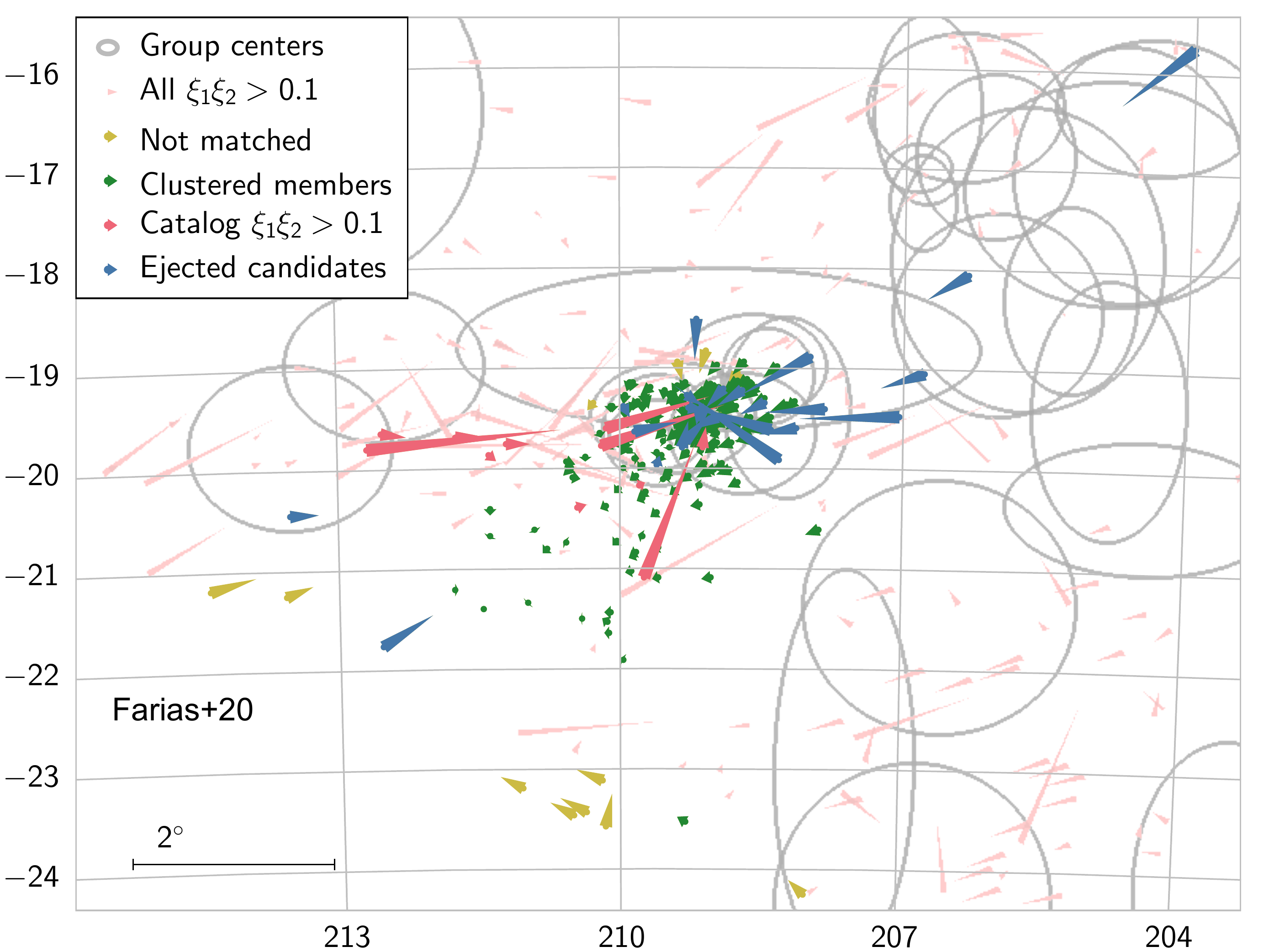}
\includegraphics[width=\columnwidth]{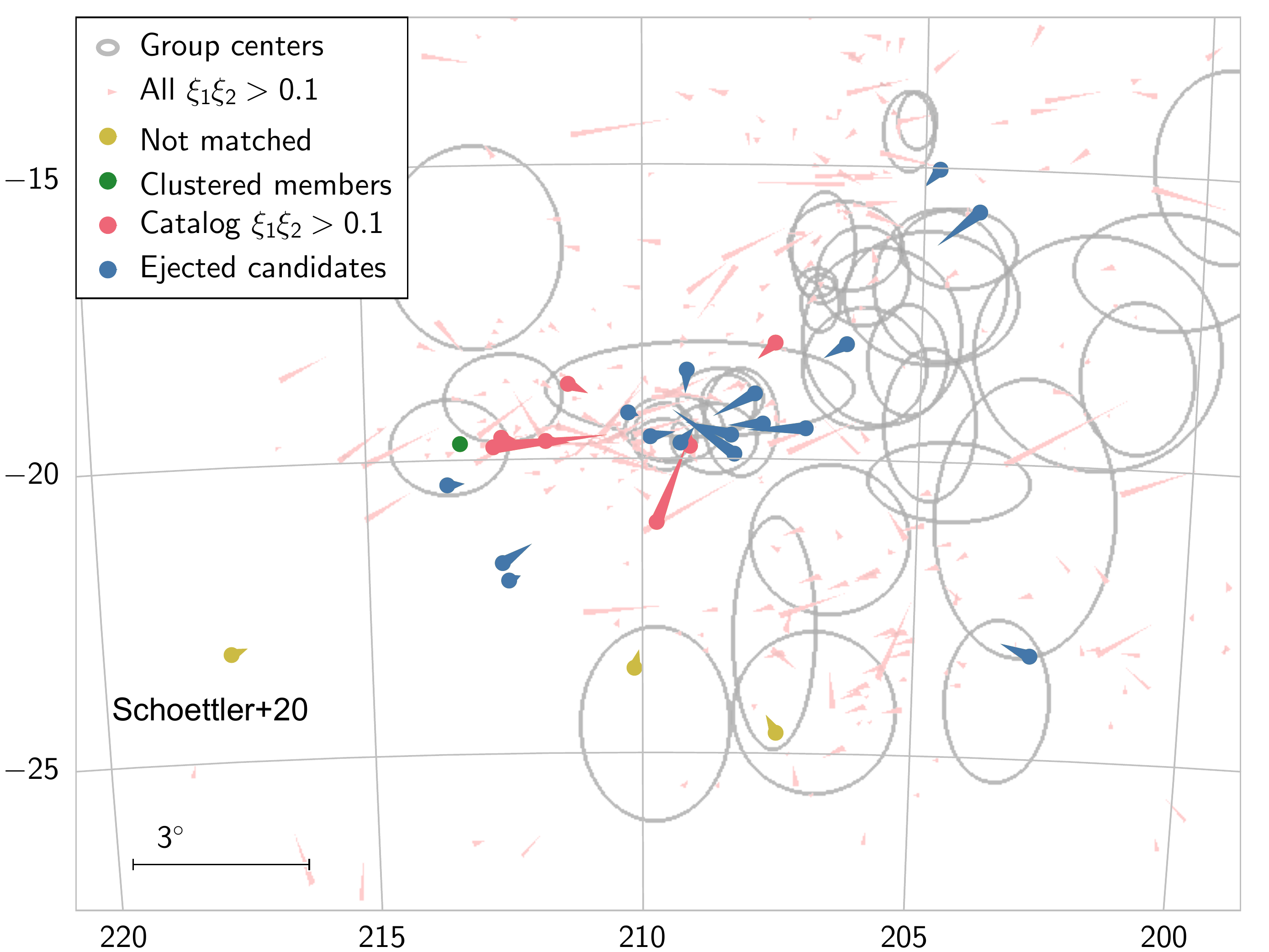}
\includegraphics[width=\columnwidth]{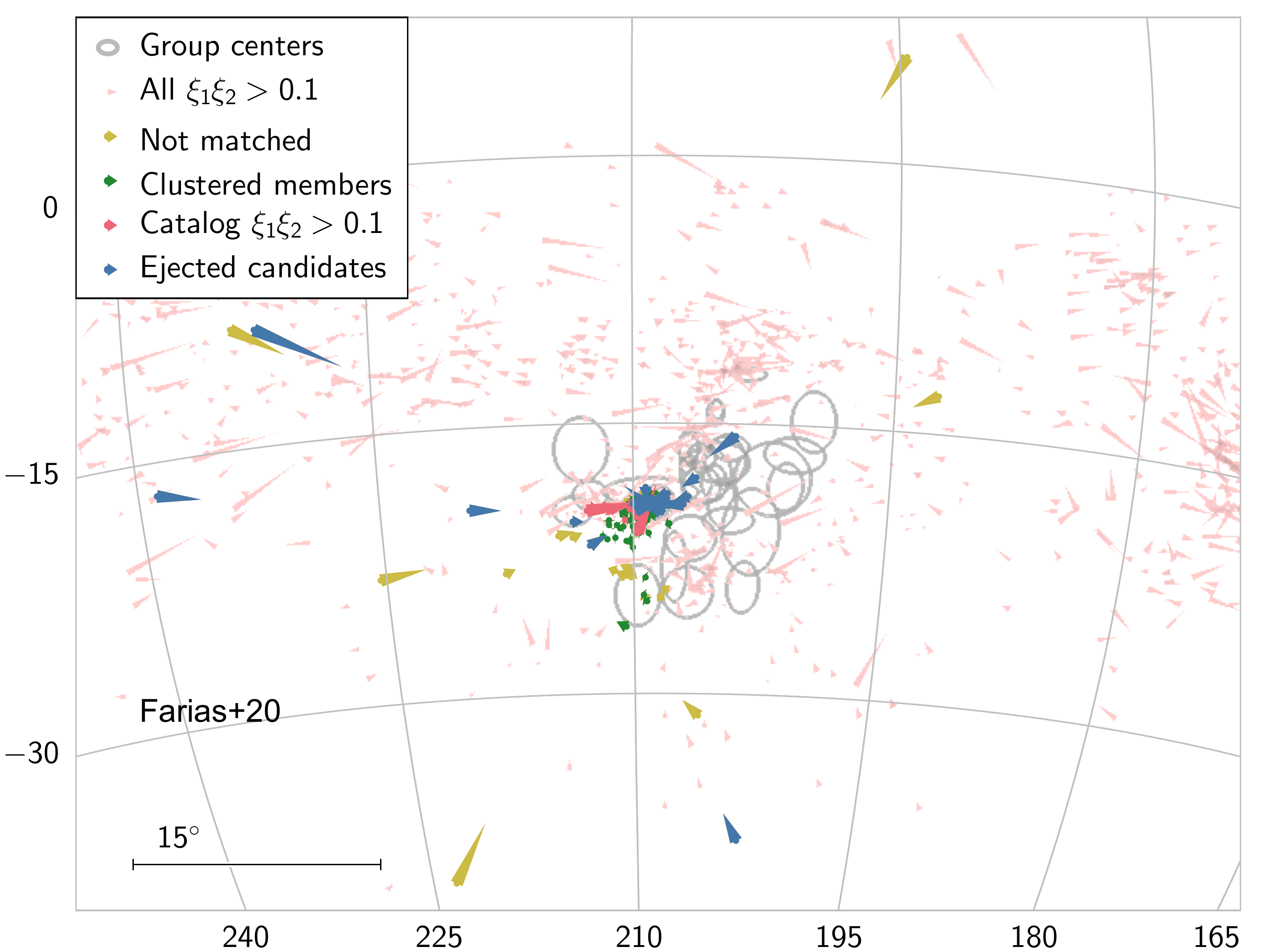}
\caption{A comparison of the trajectories of the ejected candidates in the works of \citet{mcbride2019}, \citet{schoettler2020}, and \citet{farias2020} that can be matched to the base catalog of PMS stars. Sources are color-coded based on whether the stars have met the thresholds to be selected as runaway/walkaway stars that we use in this work. Pale pink vectors show the trajectories of all sources that have $\xi_1\xi_2>0.1$ presented here, to show the dominant velocity currents in the field when evaluating the identified candidates in each catalog. Grey circles show the area on the sky corresponding to 3$\sigma_{lb,g}$ for each of the sub-groups in Orion. The plots are shown in the Galactic coordinates.
\label{fig:comparison}}
\end{figure*}

Previously, \citet{mcbride2019}, \citet{schoettler2020}, and \citet{farias2020} have searched for stars ejected from the Orion Nebula. We compare the identified candidates from these works to the catalog we derive here (Figure \ref{fig:comparison}).

\citet{mcbride2019} have identified 26 stars among known members of the ONC that have high proper motions. Of these, due to limits in color imposed by Sagitta (typically, the recovered YSO candidates have $G_{BP}-G_{RP}>2$ mag), and due to the quality of photometry that has been required in the creation of the base catalog, only 5 of these 26 sources are included in the analysis. Of these 5 sources, 2 appeared to trace back to the center of the ONC, two appeared to trace back away from the ONC as visitors to other clusters in the Orion Complex, and one had its origin unknown. In total, only one of them (a visitor, Gaia EDR3 3017044689550345856) does not meet the initial set of criteria due to a slight mismatch in age of a star ($\sim$0.5 Myr) and the traceback time ($\sim$0.6 Myr) -- we note that the uncertainty in age is not considered for the purpose of this exercise. The remaining 4 stars do meet the initial selection criteria and indeed can be traced back to various parts of Orion.

\citet{schoettler2020} have identified 85 candidates that can be traced to the ONC from up to 100 pc away from it. Of these only 27 can be matched to the base catalog from \citet{mcbride2021}, again, usually due to the color limits. Of these 27 stars, one has been selected as a core member of the Orion A molecular cloud. 3 stars do not meet the traceback criteria, due to the traceback time being longer than the age of a star and/or group. 7 stars have $\xi_1\xi_2>0.1$, and, by the strictest set of criteria, 16 sources can be recovered as runaway/walkaway candidates.

Finally, \citet{farias2020} have identified $\sim$17,000 candidates. The sources have been initially selected using a variety of different tracers of youth, including infrared excess, optical variability, position on the HR diagram, and others. Some of these criteria have a significant degree of contamination from the more evolved stars, as such, only 331 can be cross-matched against our base catalog. Of these, 267 can be recovered as core members of Orion. 22 stars do not meet traceback criteria, commonly because the conversion to the local standard of rest reference frame has significantly affected their trajectory in comparison to the heliocentric reference frame due to the large separation in distance between a star and the cluster, some due to the traceback time exceeding the age of a star and/or cluster. Of the remaining 42 stars, 11 have $\xi_1\xi_2>0.1$, and 31 meet the strictest set of criteria as walkaway/runaway candidates.

In total, we identify 249 ejected star candidates with $\xi_1\xi_2<0.1$ that can be traced back to the Orion Complex, of which 117 appear to originate from the ONC. Of these, 36 and 11 stars respectively have tangential velocity $>$30 \kms, and they can be considered as runaways.

\subsection{Statistics}

\begin{figure*}
\includegraphics[width=\columnwidth]{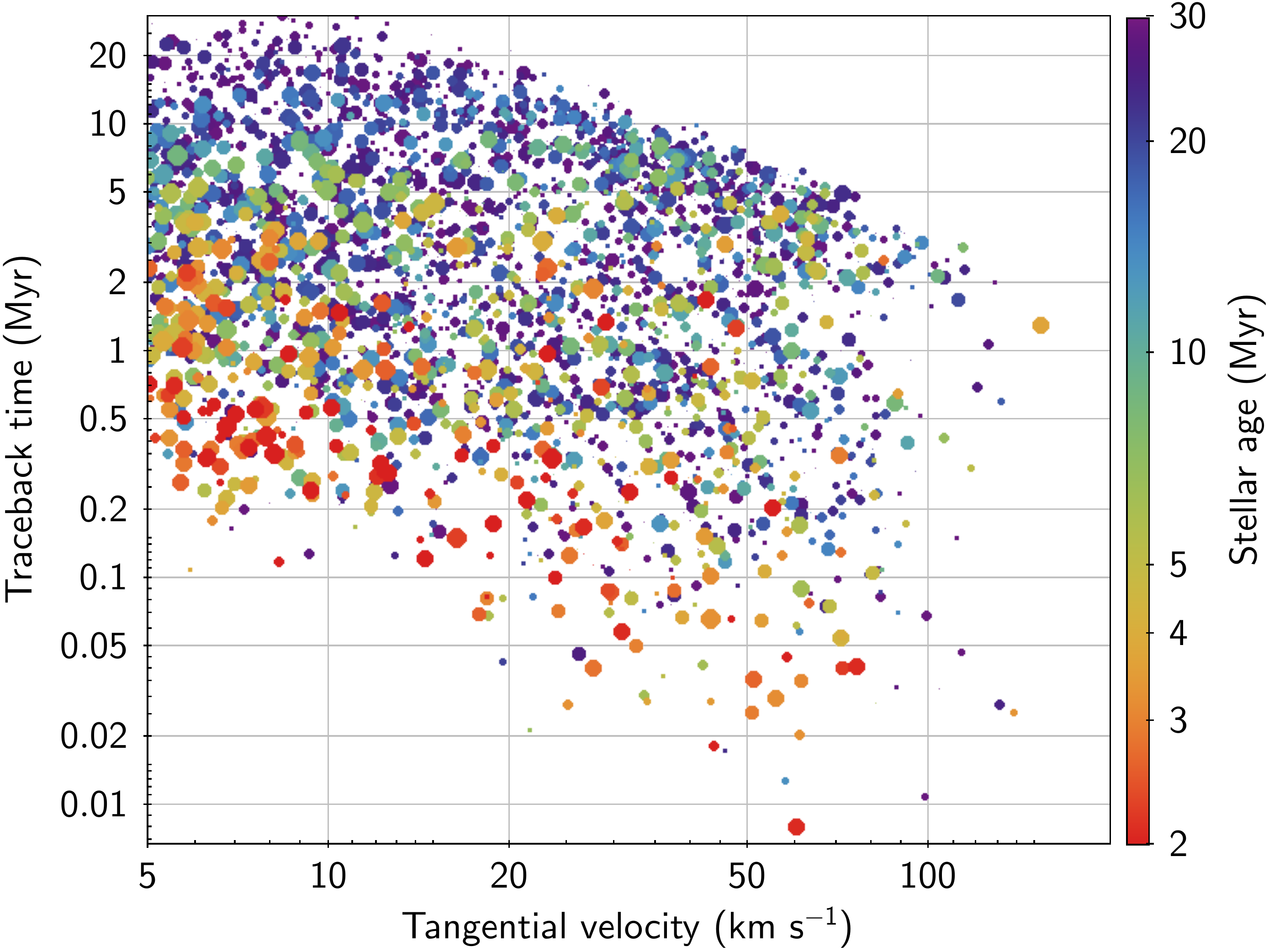}
\includegraphics[width=\columnwidth]{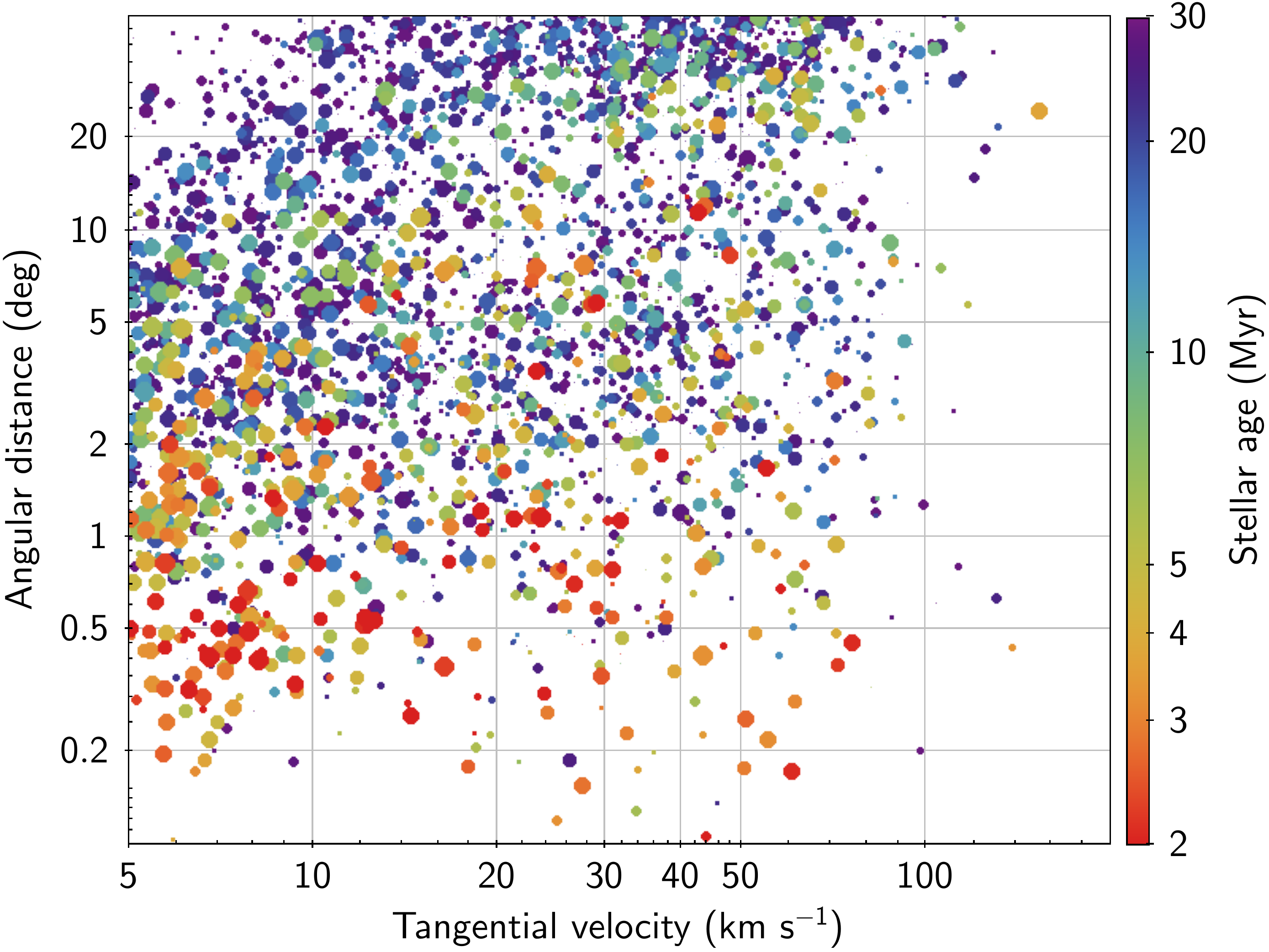}
\includegraphics[width=\columnwidth]{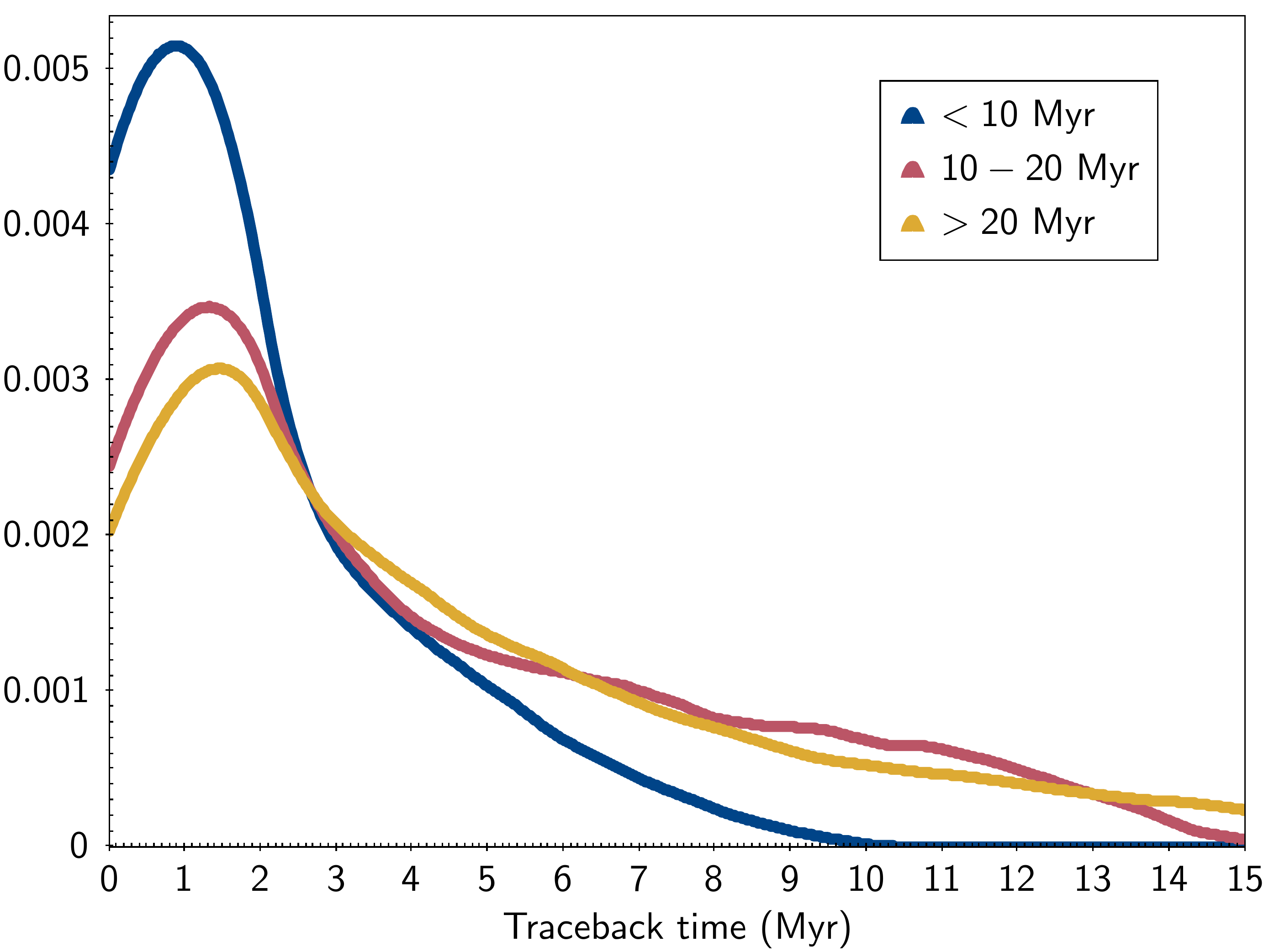}
\includegraphics[width=\columnwidth]{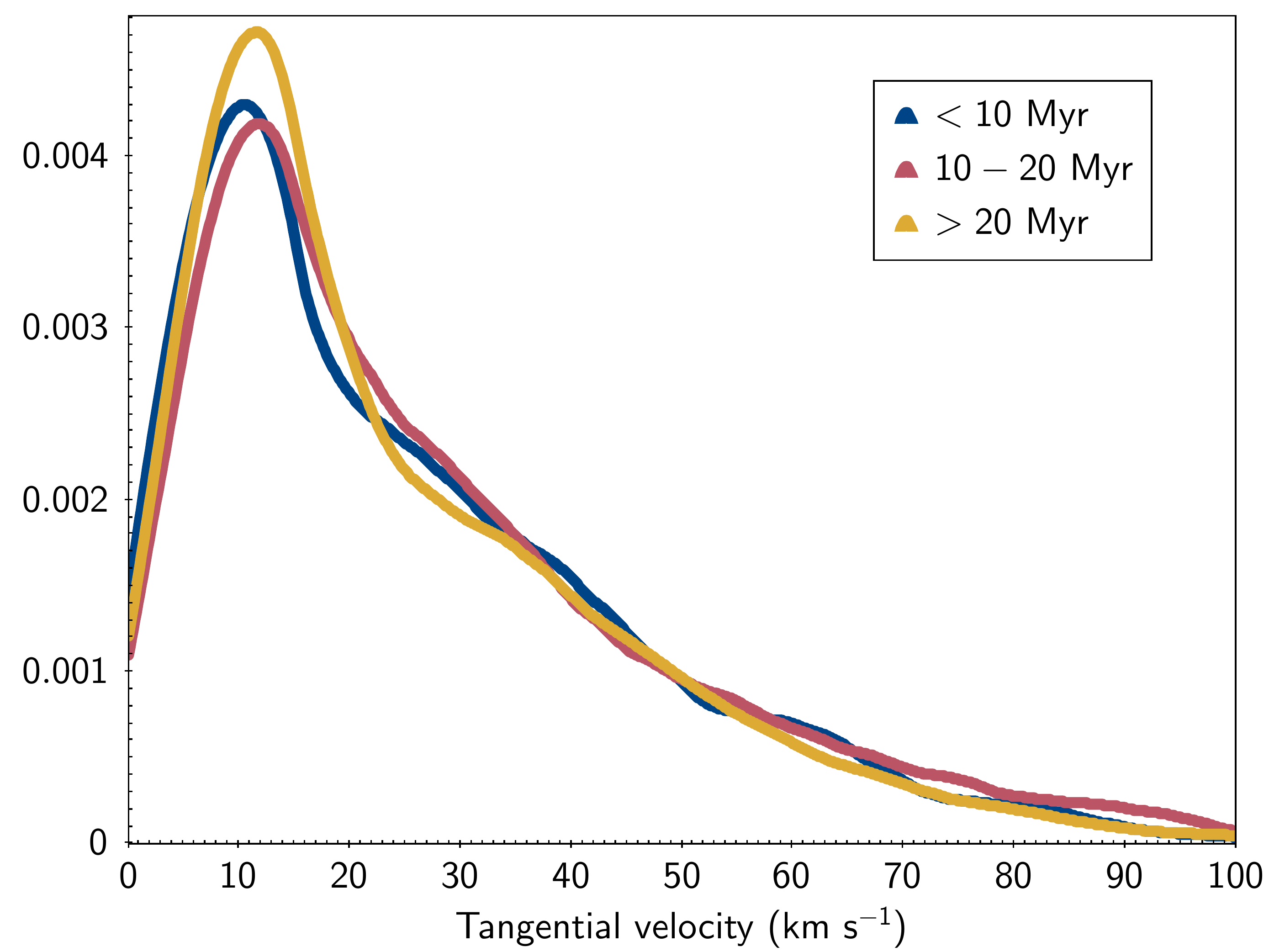}
\caption{Top: correlation between the projected tangential velocity, traceback time, and the angular distance for the candidate ejected stars with $\xi_1\xi_2<0.1$, color-coded by their age. The size of the symbol corresponds to the probability of a star being PMS, with the larger circles corresponding to the higher confidence sources, and smaller circles having a greater degree of contamination from the field stars.
Bottom: kernel density estimate showing a distribution of tangential velocity and traceback time for the sample, separated into three different age bins.
\label{fig:stats}}
\end{figure*}

We examine the properties of the candidate ejected stars in Figure \ref{fig:stats}. Across all ages of the stars, the candidates that the selection is most sensitive to are the sources that appear to have been ejected recently, within the last 2 Myr. Older stars have a longer tail in the distribution of their traceback times, as the sources that have been ejected early in their formation would have more time to travel further outward, but they may be more difficult to detect with the methodology used here.

The overall projected tangential velocity distribution of the identified candidates does not appear to vary significantly between the older and the younger stars. As such, the angular distance between the star and its projected point of origin correlates strongly with their tangential velocity, with the bulk of the sources being within 10 degrees of their parent group, and the older sources having a slightly more pronounced excess of sources at larger distances than the younger sources. 

Restricting the selection to the runaways with the relative tangential speeds $>$30 \kms\ favors the sample with more recent ejections, as they would quickly disperse away from their parent population, and they may be more difficult to recover without a traceback that is fully in 3d, since the approximation of the spherical geometry may not be sufficiently precise from identifying them at larger distances.

\subsection{Ejected pairs}

\subsubsection{Selection}

\begin{figure*}
\includegraphics[width=\textwidth]{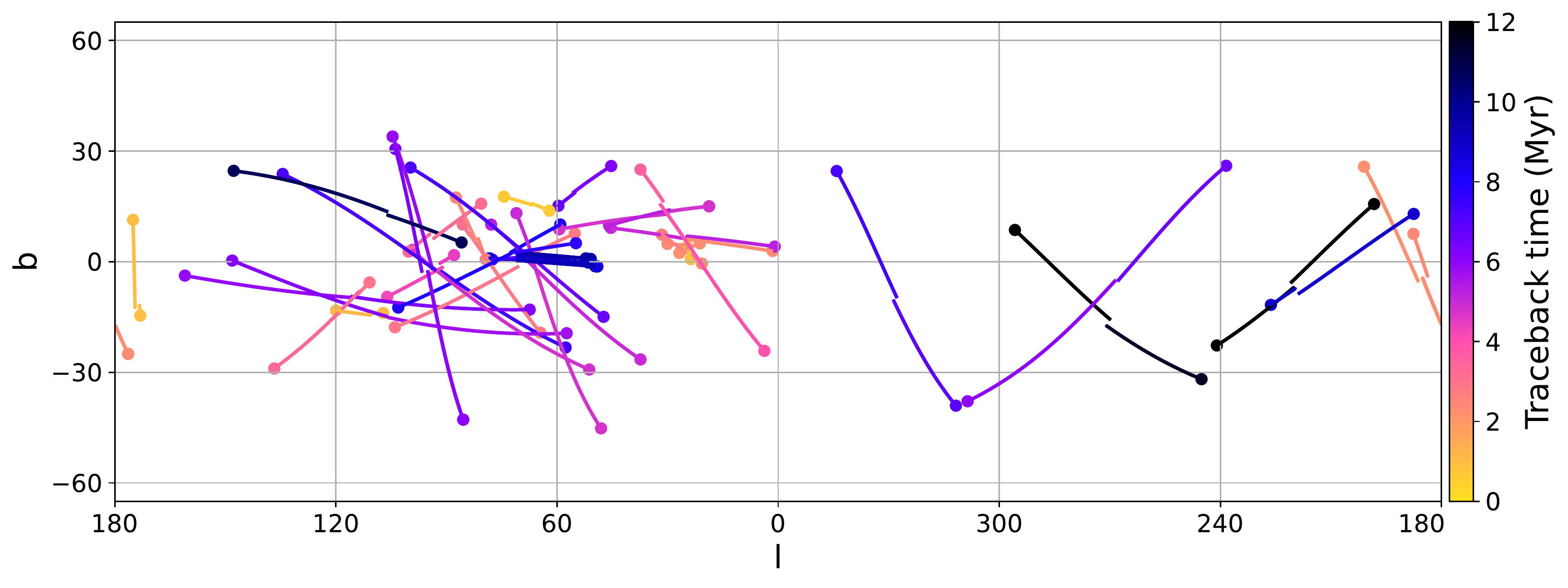}
\includegraphics[width=\textwidth]{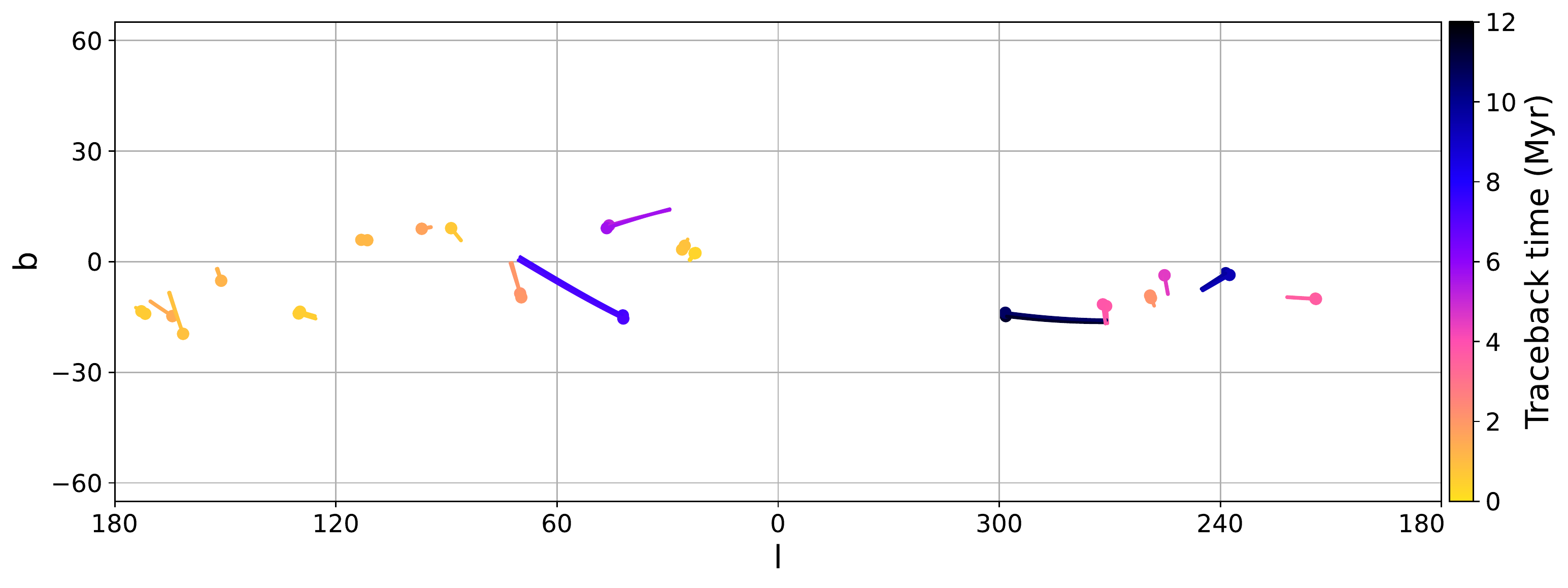}
\caption{Top: spatial distribution of the candidate interacting pairs of stars that appear to have been ejected concurrently from the same location, travelling in opposite directions. Bottom: candidate wide binary systems that may have formed in the process of the ejection (note, due to their proximity, vectors for both stars may overlap). The current position of each star is highlighted by a large circle. The vectors show their apparent path from the estimated $l_o$ and $b_o$ to their current position (signified by larger dot), they are color-coded by the traceback travel time.
\label{fig:pairs}}
\end{figure*}

In the process of disrupting a multiple system, all the stars will undergo acceleration, the precise magnitude of which depends on both the initial configuration of their orbits as well as the masses of all stars. As such it is possible to eject two stars simultaneously,
%\textcolor{red}{NL: This is actually extremely uncommon.  It is more common that you eject two stars, but separated in time.  Getting them to eah travel in opposing directions makes it a very low probability as well.  This just needs a little rephrasing.  How about:  "Binary-binary interactions are dominant for binary fractions $\gtrsim$ 10\% \citep{leigh11}, as is the case for most open clusters.  The break-up of such systems not uncommonly produces two single stars, often traveling in opposing directions \citep{leigh16}."}
each travelling away from their parent population in opposite directions from each other, similarly to AE Aur and $\mu$ Col \citep{blaauw1961,hoogerwerf2001}. %\textcolor{red}{NL:  This is most likely the result of a 4-body disintegration, which we showed in Ryu et al. 2014 or 2015, I think (I can send the ref if need, numerical scattering of N=4 in a background potential is close to the title), since if I remember right there is a recoiled binary associated with it.  You have two ejection events here, so it makes it much more convoluted and degenreat to tract back to ICs.}

For most ejected stars, finding their precise pair may be a difficult task, as most such encounters result in low ejection velocities \citep{reipurth2010}, not easily distinguishable from their parent association. In the process of mixing with their neighbors, the initial trajectory of the sources with low ejection velocity may be erased. This may be particularly common for systems with low mass ratio, in which the more massive companion is more likely to remain in a cluster, ejecting a lower mass neighbor.  %\textcolor{red}{NL:  I can provide the predicted distribution of ejection velocities for 2+1+1 ejection in the 4-body problem, if you want to compare to those.  It is in Leigh et al. 2016, on the chaotic 4-body problem in newtonian gravity.}

As such, the probability of both stars being ejected with sufficiently high velocity for them to be recovered in the catalog presented here is expected to be relatively low. Nonetheless, there do appear to be certain pairs of candidates that may have a common origin.

To identify such pairs, we require 

\begin{itemize}
    \item Using the bearing $\theta_{i}$ that has been inverted, measured from $(l_o,b_o)$ to $(l,b)$ (to prevent the distortion from a spherical geometry that may occur using $\theta$ using the current position of a star as a point of origin), the difference in $\theta_{i}$ between the pair of stars should be in the range between 175--185$^\circ$, i.e., they are moving in near-opposite directions within an arbitrarily small precision that would ensure recovery of AE Aur and $\mu$ Col.
    \item The traceback time for both stars should be within 10\% of each other, to ensure a probability of a simultaneous event, within uncertainties.
    \item The distance between their initial traceback positions $(l_o,b_o)$ is $<2^\circ$, or $<$10\% of their distance at their current positions $(l,b)$, whichever one is smaller.
    \item Although currently RVs of the candidates are in large part not available, it is possible to evaluate the distances of the candidates to reject obvious false positives. If two stars are travelling in opposite directions, and they originate from the same star forming region, the distance to their parent population has to be between the distances of both stars, within the tolerance for the uncertainties and the depth of the region.
    \item Both stars have $\xi_1\xi_2<$0.1.
\end{itemize}

\begin{figure}
\includegraphics[width=\columnwidth]{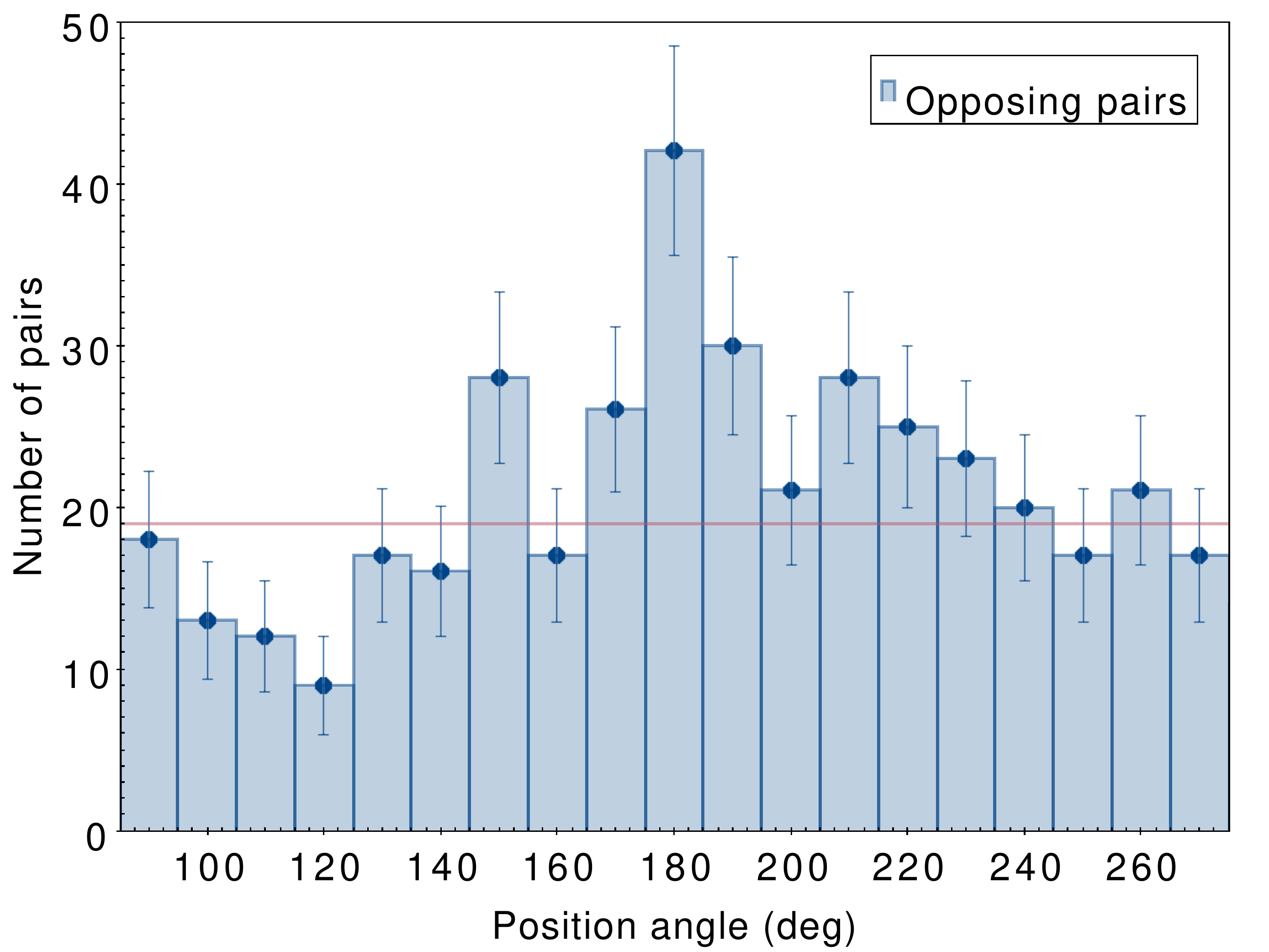}
\caption{The number of system that are selected using the criteria for the opposing pairs, but using different requirement for the angle. Typical number of systems that appear to move relative to one another at random angles, but project back to a similar origin within a similar period of time is 19 pairs (shown as red line), there is a significant excess at the angle of 180 degrees, corresponding to the stars moving in the opposite directions.
\label{fig:orientation}}
\end{figure}

In total, 42 pairs satisfy these cuts; they are shown in Figure \ref{fig:pairs}. There may be chance coincidences in this sample, consisting of pairs that only appear to move opposite of one another, and that do indeed originate from the cluster, but wouldn't necessarily trace back to the same system. However, repeating the above selection for angles other than 175--180$^\circ$ results in a typical number of 19 systems that are selected, i.e., there is a significant excess in the number of systems moving opposite of one another (Figure \ref{fig:orientation}). We further note that if we change some of the criteria -- e.g., keep all of the requirements regarding the traceback to a given cluster the same, but force the minimum separation between stars at the initial position to be $>2^\circ$, there is no preferred direction of motion between these random matches. As such, even though this sample of the opposing pairs may be contaminated by chance coincidences, we expect a subset of these systems to truly be simultaneous ejections.

Additionally, it may be possible for the two of the ejected stars to form a binary system in the process of the encounter. Compact binaries can be detected in the future through spectroscopic followup; however, in wide binaries, both stars may be fully resolved with Gaia, in which case, both stars are expected to have very similar kinematics, and have very similar trajectories. An example of such a system are Brun 259 and V1961 Ori \citep{mcbride2019}, two stars with separation of $\sim7000$ au that appear to have been concurrently ejected from the ONC 0.1 Myr ago. This system is notable, as binary systems with such wide separations are not expected to be stable in such a dense cluster. We use criteria similar to the above to identify new candidates:

\begin{itemize}
    \item $\theta$ of both stars is within 5 degrees of each other.
    \item The traceback time for both stars should be within 10\% of each other.
    \item Their projected separation is within 10 pc.
    \item Their parallaxes are consistent with 10 pc separation, within errors.
    \item Both stars have $\xi_1\xi_2<$0.1 and $\xi_1<$0.1, this cut is stricter, to further minimize any contamination from the sources that may belong to various moving groups, however distantly.
\end{itemize}

This results in a catalog of 19 candidates, shown in Figure \ref{fig:pairs}. Of these, 6 have projected separation within 1 pc. Recently, \citet{el-badry2021} have produced a catalog of wide binary candidates across the entire sky, containing $>$1 million pairs with 3d $<$1 pc. This catalog is able to recover 5 of these 6 stars. We note that cross-matching against full sample sources that meet our initial set of criteria results produces 697 stars in common, however, there is a significant contamination in this sample from various moving groups and cases of clear mismatch in age of the stars in the pair, such as pre-main sequence star and a much more evolved counterpart, which are unlikely to be true binaries. As such the selection of 19 pairs is more conservative, however there may be other systems in the sample that are missed by our selection. 

\subsubsection{Properties}

\begin{figure}

\includegraphics[width=\columnwidth]{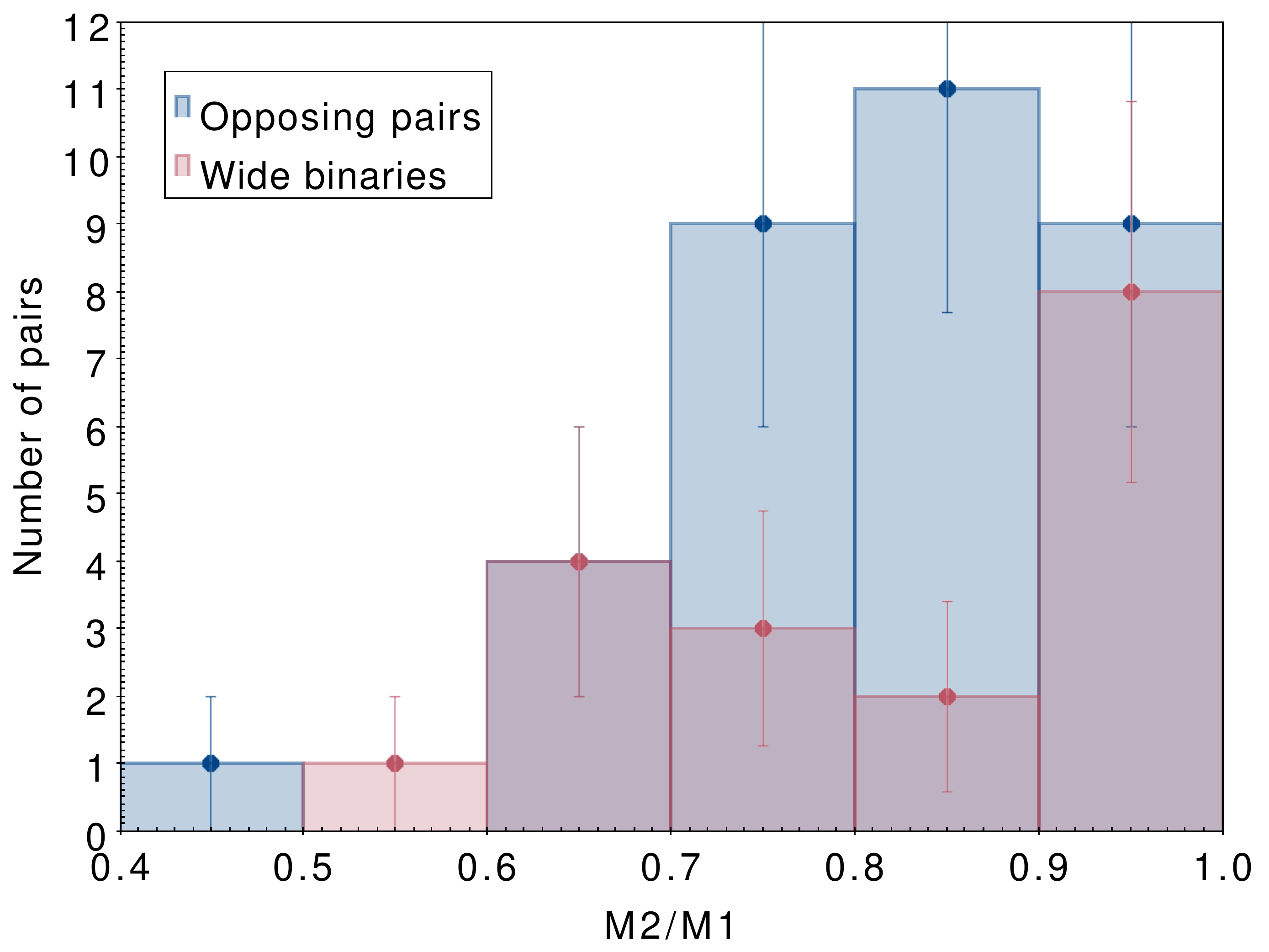}
\includegraphics[width=\columnwidth]{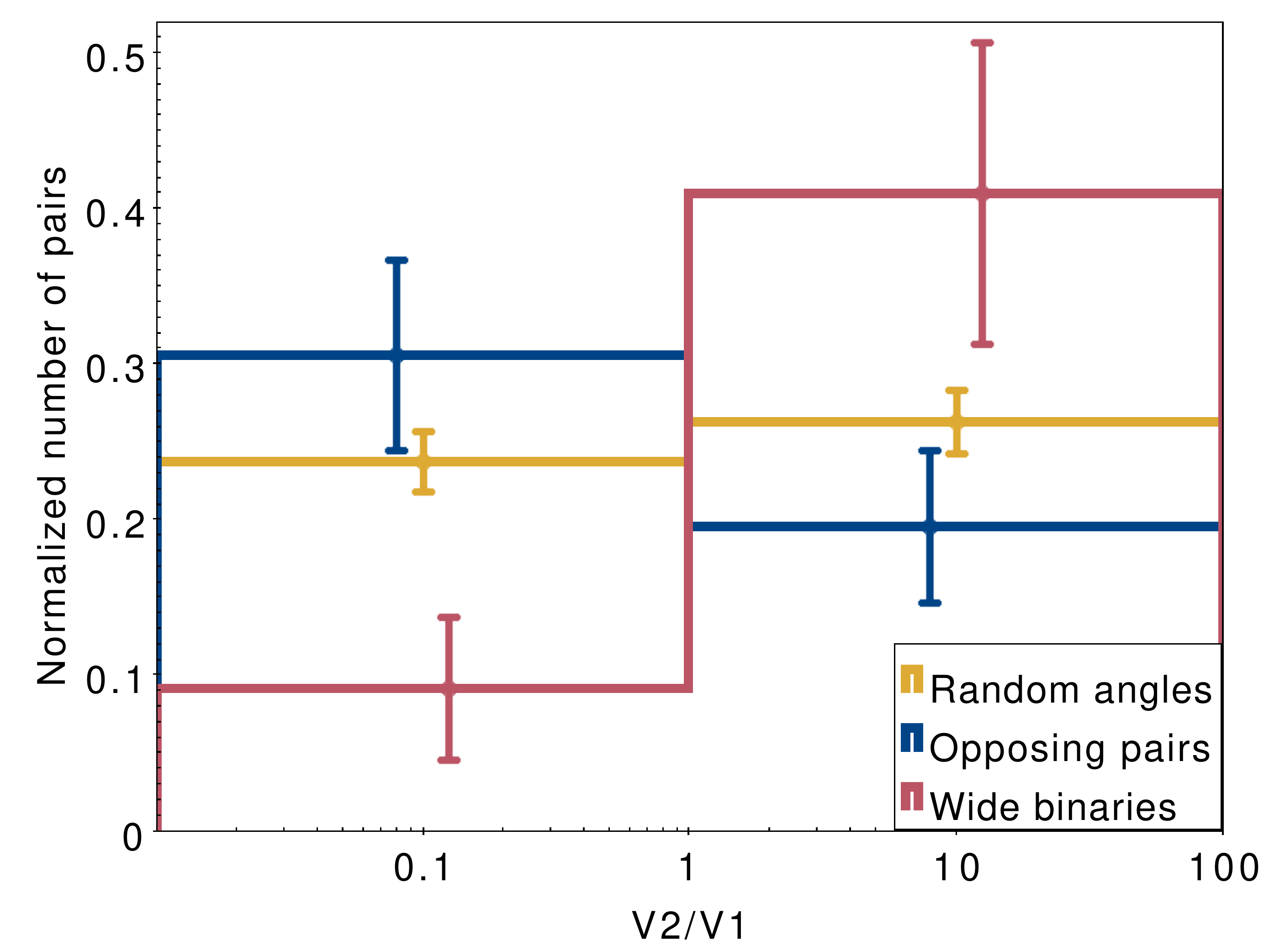}
\caption{Top: The comparison of mass ratios between the lower and higher mass star in the system of the opposing pairs, and among the wide binaries. Bottom: The comparison of the tangential velocity ratios between lower mass and higher mass star between the opposing pairs, wide binaries, as well as pairs chosen as moving at random angles relative to one another. The uncertainties are derived using sqrt(n) approximation.
\label{fig:pairsstats}}
\end{figure}
\begin{figure}
\includegraphics[width=\columnwidth]{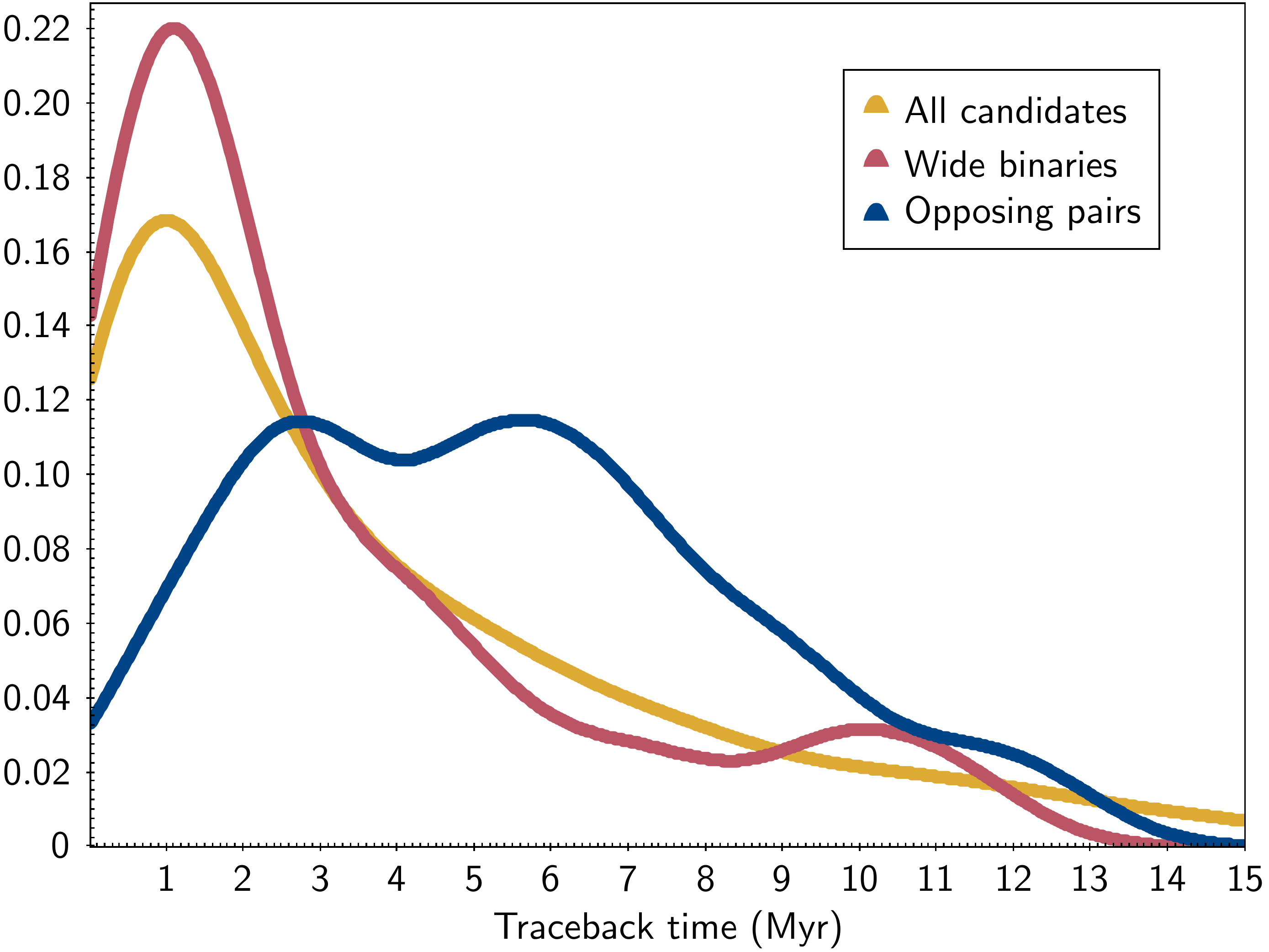}
\includegraphics[width=\columnwidth]{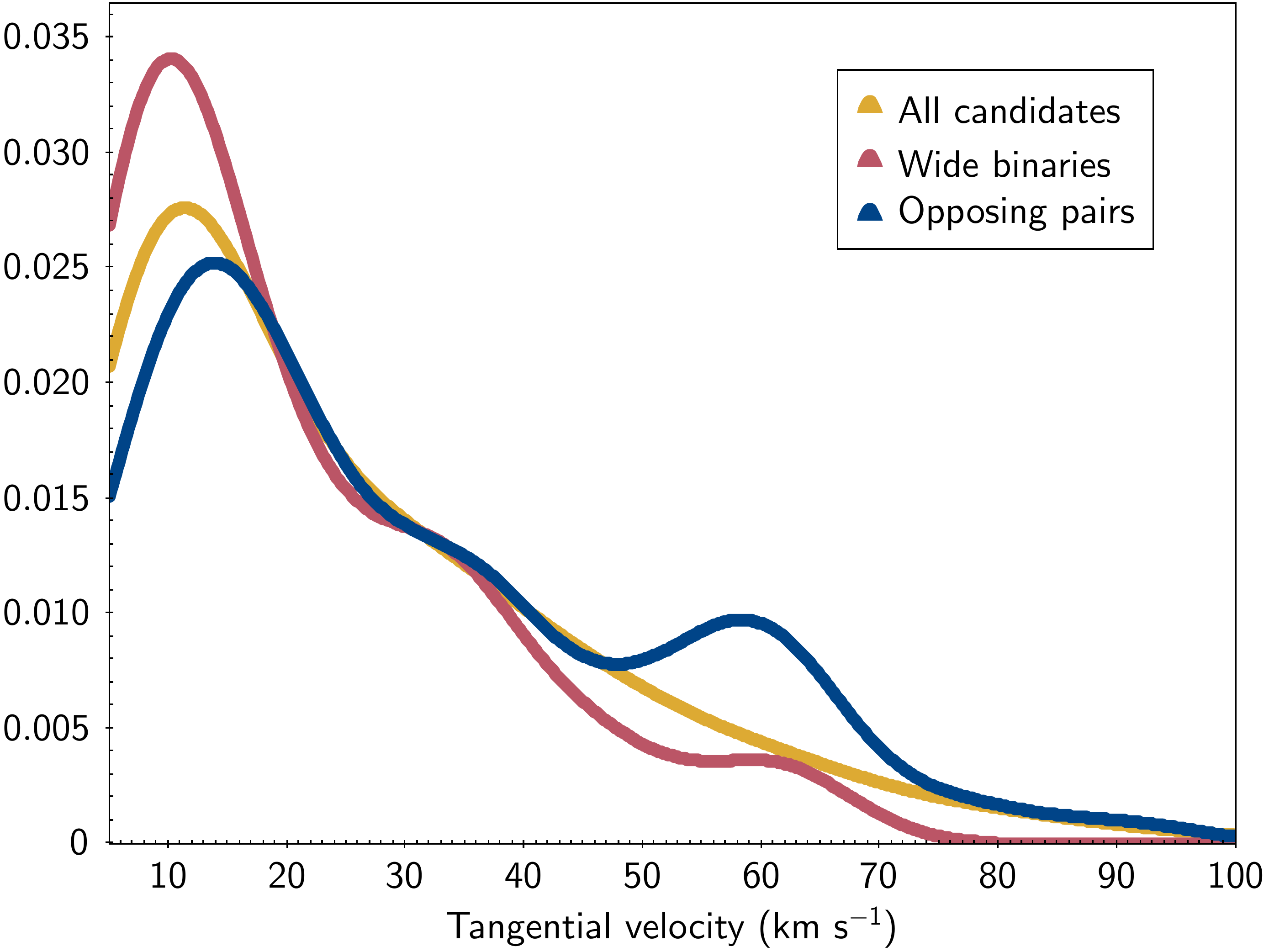}
\caption{Kernel density estimate showing a distribution of tangential velocity and traceback time for the full sample of candidate ejected stars, as well as the sources that are a part of opposing pairs and wide binaries.
\label{fig:pairsvel}}
\end{figure}

We examine the properties that these stellar pairs have. To estimate masses of the individual stars, we cross-match the catalog of the identified systems against TESS Input Catalog \citep{stassun2019}, which has an estimate of mass through photometry. We note that these young stars are moving almost completely horizontally on the HR diagram along the Hayashi tracks, as such minor discrepancies in the age of the star does not have a substantial effect on the mass.

Comparing the mass ratio $q$ within each pair, we find that the opposing pairs tend to have $q$ consistent with being uniform (with the detection bias against identifying low $q$ systems, due to the typical masses of the stars recovered in the base catalog). However, wide binaries show a preference towards unity, almost half of the systems have $q>0.9$, with a significant deficit of systems with $q$ between 0.7--0.9 in comparison to the opposing pairs (Figure \ref{fig:pairsstats}). There may be a slight trend of wide binaries to also be increasingly more common at lower $q$, but due to the small number of such systems, and due to the aforementioned detection bias it is difficult to definitively state it.

We also examine the tangential velocity ratio between the lower mass and higher mass star in the system (Figure \ref{fig:pairsstats}). We find that in wide binaries, by construction of the catalog, velocities of the pairs are very similar, however the lower mass star tends to have slightly higher speed than the higher mass one: typically $v_2/v_1\sim1.05$. This is unlikely to be caused by the orbital motion. For a system with the separation of 1 pc, the orbital period is $\sim$100 Myr, increasing to 3 Gyr for separations of 10 pc, i.e., much longer than the age of these stars. Rather, any orbit between these stars is likely to be unstable, and they are unlikely to necessarily be gravitationally bound. While they most likely have been ejected in the same event, conservation of angular momentum has imparted a slightly lower mass star greater velocity, and they will continue to grow further apart. The only reason why they can be observed as a pair is due to their youth, since their velocities have not evolved significantly post ejection. Indeed, on average, these wide binaries tend to be on average younger than the full sample of candidate ejected stars, and they tend to be marginally slower, as they are more likely to lose coherence the further they are from their parent population (Figure \ref{fig:pairsvel}).

An opposite trend is observed among the opposing pairs: as the precise location of the ejection within a region is unknown, at larger distances minute deviations in the initial angle of ejection become less apparent. As such to get further out, these opposing pairs tend to be somewhat older and somewhat faster moving than the full sample (Figure \ref{fig:pairsvel}). And, surprisingly, a more massive star tends to have a slightly faster velocity than a lower mass one (Figure \ref{fig:pairsstats}). Pairs selected through similar set of criteria, but moving relative to each other at any angle other than 180$^\circ$ tend to have a more symmetric distribution of velocities between $v_2$ and $v_1$ (Figure \ref{fig:orientation}). The excess in the opposing pairs is slight, however, and due to a small number of pairs, the statistical significance is low. 

Nonetheless, we consider possibilities of how such an excess of faster moving massive stars in the pair may arise. One of them is that initially, both stars may have belonged to an unstable quadruple system. Initially, the least massive star in the system may have been ejected, removing energy from the system in the process, making it more compact but still unstable. The second least massive star would then also be ejected shortly afterwards, and with a new configuration the resulting ejection would often \citep[e.g.][]{leigh2016,ryu2017,ryu2017a} produce slightly higher velocities, provided the virial ratio is close to zero \citep{leigh2016}. However, a problem with such a scenario is that an ejection would not be simultaneous, but rather two separate ejections in a relatively rapid succession of each other, each launched in a random direction. Such systems may be present among the systems that meet the criteria for the opposing pairs but have different angles of motion; future follow up may help to more definitively identify such systems, separating them from chance coincidences, enabling a way to trace back their formation history. However, while this scenario can contribute to a ``floor'' in the number of pairs across all orientations (Figure \ref{fig:orientation}), this cannot explain the preference for the pairs to move 180$^\circ$ away from one another.

Another method would be possible through a disintegration of a triple system: through ejecting one of the stars, the newly formed binary system may have received enough recoil to also have been ejected -- both this system and the ejected stars would be moving 180$^\circ$ relative to one another. And, the binary system would most likely be more massive than the ejected single, even though individual stars in that binary may be of lower mass, which could be responsible for the observed velocity signature.

In large part, a companion with separations $<$0.7'' is less likely to be present as a part of the Gaia catalog. Although Gaia DR3 has produced various catalogs of stellar multiplicity \citep{gaia-collaboration2022}, including astrometric, spectroscopic, and eclipsing binaries, they are highly incomplete (especially given the aforementioned issues with RVs in the sample of these young stars). As such, a future search for closer companions among these opposing pairs may better enable understanding their formation.

\section{Conclusions} \label{sec:conclusions}

In this work we present a first homogeneous search for candidate young runaway/walkaway stars across the entire sky within the Solar Neighborhood, resulting in a catalog of 3354 sources. These candidates were selected from a photometric catalog of pre-main sequence stars that can be separated from the field stars with a relatively high degree of confidence. These sources can be traced back to the nearby star forming regions within the lifetime of a star and their parent association, most commonly within a few Myr. They are also distinct from the dominant velocity currents of dissolving young moving groups that are found in their vicinity.

As part of these analysis, we also identify candidate interacting pairs that appear to have been ejected concurrently in the same event, of which 42 appear to be travelling in opposite directions, and 19 appear to have formed a wide binary in the process of their ejection.

The traceback is performed solely in the plane of the sky, as currently only sparse data are available pertaining to the radial velocity information for these stars. Despite the wealth of recently released Gaia DR3 wealth of RVs across the Solar Neighborhood, these data offer poor constraints on the radial velocity of low mass stars younger than 100 Myr. 

As such these candidates require spectroscopic follow-up observations to enable the unequivocal confirmation of both their youth that would complement the initial photometric selection, as well as their status as walkaway/runaways through a full 3d traceback. Nonetheless, once confirmed, these candidates can offer insight on the initial dynamical conditions within their parent associations that have led to these stars being ejected.

\section*{Acknowledgements}

This work has made use of data from the European Space Agency (ESA)
mission {\it Gaia} (\url{https://www.cosmos.esa.int/gaia}), processed by
the {\it Gaia} Data Processing and Analysis Consortium (DPAC,
\url{https://www.cosmos.esa.int/web/gaia/dpac/consortium}). Funding
for the DPAC has been provided by national institutions, in particular
the institutions participating in the {\it Gaia} Multilateral Agreement.  NWCL gratefully acknowledges the generous support of a Fondecyt Iniciaci\'on grant 11180005, as well as support from Millenium Nucleus NCN19-058 (TITANs) and funding via the BASAL Centro de Excelencia en Astrofisica y Tecnologias Afines (CATA) grant PFB-06/2007.  NWCL also thanks support from ANID BASAL project ACE210002 and ANID BASAL projects ACE210002 and FB210003.
%%%%%%%%%%%%%%%%%%%%%%%%%%%%%%%%%%%%%%%%%%%%%%%%%%
\section*{Data Availability}

This work uses the data previously made available in \citet{mcbride2021}. The full tables are included in the Supplemental Materials and will be made available on Vizier.



\bibliographystyle{mnras}
\bibliography{main.bbl} % if your bibtex file is called example.bib

%\begin{thebibliography}{99}
%\end{thebibliography}

%%%%%%%%%%%%%%%%%%%%%%%%%%%%%%%%%%%%%%%%%%%%%%%%%%

%%%%%%%%%%%%%%%%% APPENDICES %%%%%%%%%%%%%%%%%%%%%

%%%%%%%%%%%%%%%%%%%%%%%%%%%%%%%%%%%%%%%%%%%%%%%%%%

% Don't change these lines
\bsp	% typesetting comment
\label{lastpage}
\end{document}